\documentclass[lettersize,journal]{IEEEtran}
\usepackage{amsmath,amsfonts}
\usepackage{algorithmic}
\usepackage{algorithm}

\usepackage{array}
\usepackage[caption=false,font=footnotesize,labelfont=rm,textfont=rm]{subfig}
\usepackage{textcomp}
\usepackage{stfloats}
\usepackage{url}
\usepackage{verbatim}
\usepackage{graphicx}
\usepackage{cite}
\usepackage{multirow}
\usepackage{xcolor}
\newcommand{\bluesection}[1]{{\color{black}#1}}
\newcommand{\blue}[1]{{\color{black}#1}}
\newcommand{\purple}[1]{{\color{black}#1}}

\begin{document}

\title{iBA: Backdoor Attack on 3D Point Cloud via Reconstructing Itself }

\author{Yuhao Bian, Shengjing Tian, Xiuping Liu
        \thanks{https://ieeexplore.ieee.org/document/10659845}


       }



\maketitle

\begin{abstract}
The widespread deployment of deep neural networks (DNNs) for 3D point cloud processing contrasts sharply with their \bluesection{vulnerability} to security breaches, particularly backdoor attacks. \bluesection{Studying these attacks is crucial for raising security awareness and mitigating potential risks. However, the irregularity of 3D data and the heterogeneity of 3D DNNs pose unique challenges. Existing methods \purple{frequently} fail against basic point cloud preprocessing or require intricate manual design.} Exploring simple, imperceptible, effective, and difficult-to-defend triggers in 3D point clouds remains challenging. \bluesection{To address this issue, we propose iBA, a novel solution utilizing a folding-based auto-encoder (AE). By leveraging united reconstruction losses, iBA enhances both effectiveness and imperceptibility. Its data-driven nature eliminates the need for complex manual design, while the AE core imparts significant nonlinearity and sample specificity to the trigger, rendering traditional preprocessing techniques ineffective. Additionally, a trigger smoothing module based on spherical harmonic transformation (SHT) allows for controllable intensity. We also discuss potential countermeasures and the possibility of physical deployment for iBA as an extensive reference. }
Both quantitative and qualitative results demonstrate the effectiveness of our method, achieving state-of-the-art attack success rates (ASR) across \bluesection{a variety of victim models, even with defensive measures in place. iBA's imperceptibility is validated with multiple metrics as well}.

\end{abstract}


\section{Introduction}
\begin{figure} 
    \centering
  \subfloat[\label{display_1}]{%
       \includegraphics[width=0.3\linewidth]{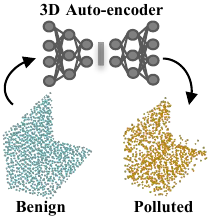}}
    \hfill
  \subfloat[\label{display_2}]{%
        \includegraphics[width=0.62\linewidth]{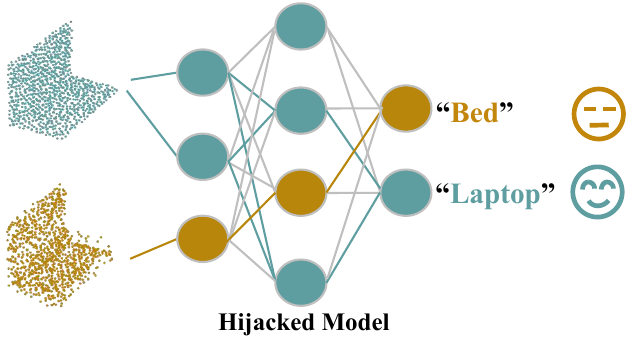}}
  \caption{The sketch of our work. (a) \bluesection{We design a sample-specific and imperceptible backdoor attack on 3D point clouds with a folding-based AE}; (b) The victim model will be hijacked and manipulated once trained on the attacked dataset.} 
  \label{display} 
\end{figure}

\IEEEPARstart{D}{eep} learning for 3D applications has witnessed significant advancements in recent years, driven by the growing spectrum of application demands and the widespread availability of sensing technologies. The point cloud, owing to its versatility and geometry-sensitive characteristics, has become a preferred medium for 3D object representation. A diverse array of DNNs tailored for 3D point clouds\cite{pointnet}\cite{pointnet++}\cite{dgcnn}\cite{pointcnn}\cite{pct} have emerged, positioning themselves as foundational technologies in sectors such as autonomous driving\cite{driving-tian}, healthcare\cite{health-liu}, and augmented reality\cite{reality-kastner}.
Meanwhile, backdoor attacks\cite{badnets} \cite{blended} \cite{sig} \cite{wanet} \cite{inputaware} \cite{ssba} \cite{pointba} \cite{irba} \cite{mba} are posing an increasingly serious threat to these DNNs. Backdoor attackers embed a trigger in part of the training data while changing the labels to target labels before sharing the contaminated dataset \purple{with} the community. Once users download and train their DNNs on these third-party datasets, a \textit{backdoor} is opened in the DNNs. They will work normally under regular circumstances. However, when encountering triggered data Fig.\ref{display_2}, they will infer incorrectly according to the attacker's predetermined plan. For those security-critical applications, the imperceptible and destructive nature of backdoor attacks appears particularly lethal, which has garnered widespread attention from researchers in recent years.

Although backdoor attack algorithms are rapidly evolving, the vast majority focus on the image domain\cite{badnets}\cite{blended}\cite{sig}\cite{wanet}\cite{inputaware}\cite{ssba}, \bluesection{while} those targeting 3D point clouds are still scarce. 
Unlike the regularity of 2D images, 3D point clouds are inherently point sets discretely sampled from underlying surfaces. The sparse and irregular nature of point clouds introduces unique challenges and complexities \bluesection{in designing backdoor attacks}. PointBA-I\cite{pointba} and PCBA\cite{pcba} treat point clouds as sets, integrating a point cluster as the trigger near the point cloud. While they often achieve high attack success rates (ASRs), their triggers are vulnerable to statistical methods like statistical outlier removal (SOR)\cite{sor}. \bluesection{In contrast}, PointBA-O\cite{pointba} and NRBdoor\cite{nrbdoor} view point clouds as discrete surfaces, embedding triggers through rotational transformations in \bluesection{3D} space. However, rotational data augmentation, a common practice for training 3D point cloud networks, can easily nullify such attacks. 
In response, Gao et al.\cite{irba} proposed IRBA, \bluesection{which employs} Gaussian smoothing over multiple local linear transformations. This sample-specific, covert trigger's nonlinear nature effectively resists various data augmentation techniques. Yet, similar to WaNet\cite{wanet} in images, the success of IRBA relies on deliberate manual design. However, the irregularity of 3D point clouds increases the difficulty of designing. Exploring simple, effective, imperceptible, and difficult-to-defend triggers in 3D point clouds remains challenging.

\bluesection{
Inspired by the intuition that 3D AEs can introduce special geometric perturbations during the reconstruction and motivated by \purple{pioneers} in 2D\cite{inputaware}\cite{ssba}, we now seek a data-driven solution. However, the reconstruction error in 3D point clouds cannot directly serve as a backdoor trigger. There are two key challenges to be overcome. 
The first challenge is the significant heterogeneity of 3D victim models. The design of 3D models varies, and their ability to capture geometric features also differs. \blue{Therefore, an attack can be biased unless its trigger is highly adaptable.} For instance, the diffusion-based trigger is effective against PointNet++ and DGCNN but fails against PointNet, while the high-frequency point jitter can deceive PointNet++ but leave PointNet and DGCNN unaffected. 
\blue{Both sufficient attack intensity and diverse geometric perturbations can help tackle this challenge. Thus, we implement our attack, iBA, with a folding-based AE\cite{foldingnet} (see Fig.\ref{display_1}), which reconstructs an arbitrary point cloud from a fixed 2D grid. This inductive bias endows the generated shapes with special fingerprints, such as grid-like textures, distortions, and noises\cite{foldingnet}. The fingerprints can provide adequate reconstruction error and rich geometric perturbations across a broad frequency range (see Sec.\ref{pattern}). Moreover, we introduce an additional distribution-oriented loss\cite{wd} to optimize the reconstruction error and enhance the imperceptibility, making it more suitable as a backdoor trigger.}
The second challenge lies in controlling the intensity of the reconstruction error. The image reconstruction exhibits natural pixel-wise correspondence, allowing the attacker to adjust the trigger intensity through linear interpolation \purple{to balance the attack imperceptibility and effectiveness}. Nevertheless, the reconstructed 3D point cloud loses point-wise correspondence with the benign source, rendering such trivial intensity adjustment infeasible. To mitigate this, we design a trigger smoothing module based on spherical harmonic transformation (SHT) to approximate a smooth deformation from the benign point cloud to the poisoned one, which improves the flexibility of iBA and expands its applicable scenarios. This approach can also serve as a reference for other attacks where correspondence is lost.

iBA subtly transforms \purple{the} imperfections in point cloud reconstruction into a lucrative tool for backdoor attacks, \bluesection{offering} three-fold advantages. Firstly, its data-driven core, based on the reconstruction loss, \bluesection{automatically enhances the trigger's imperceptibility and eliminates the need for intricate manual design}. Secondly, the deep generative network naturally provides the trigger with special geometric perturbations and high nonlinearity, fortifying it against common data augmentations. Lastly, the trigger is sample-specific, as the trigger patterns are conditioned on the benign source, raising the defense threshold. 

Experimental results in Sec.\ref{experiments} verify the effectiveness of our approach across diverse 3D point cloud classifiers. Notably, our method achieves a significantly elevated ASR while maintaining enhanced imperceptibility. Our technique also demonstrates robust resistance to a broad spectrum of data augmentation strategies. 
In addition, we discuss the pattern of our trigger, countermeasures under several user-informed scenarios, and the potential for physical deployment in Sec.\ref{pattern},  Sec.\ref{potential_defense}, and Sec.\ref{physical_potential}, respectively, to provide a broader perspective. }

Our contributions can be summarized as follows:
\begin{itemize}
\item{We propose iBA, a novel poisoned-label backdoor attack for 3D point clouds, which automatically generates sample-specific, effective, and imperceptible trigger patterns through a \bluesection{folding-based} AE. }
\item{We design a trigger smoothing module based on SHT, which enables a continuous transformation from \bluesection{benign data to backdoor data}, allowing for a controllable attack. }
\item{\bluesection{iBA} demonstrates significant performance in terms of imperceptibility and effectiveness. It also achieves state-of-the-art ability} to \purple{penetrate multiple defenses}.

\end{itemize}

\begin{table}[H]
\begin{center}
\caption{We roughly categorize three representative point cloud backdoor attack methods and our iBA based on four attributes} 
\label{tab:types}
\renewcommand \arraystretch{1.2}
\resizebox{\linewidth}{!}{
\begin{tabular}{c|c|c|c|c}
\hline
                & PointBA-I  & PointBA-O  & IRBA       & iBA (Ours)\\ \hline \hline
Interaction     & \checkmark &            &            &             \\ \hline
Transformation  &            & \checkmark & \checkmark & \checkmark  \\ \hline
Sample-specific &            & \bluesection{--}      & \checkmark & \checkmark  \\ \hline
Data-driven     &            &            &            & \checkmark  \\ \hline

\end{tabular}
}
\end{center}
\end{table}

\section{Related Work}

\subsection{DNNs for 3D Point Clouds} \label{3d_victims}

From the perspective of backdoor attacks, DNNs for 3D point clouds are primarily studied as the \textit{victim models}.
Currently, numerous high-performance \bluesection{models have been introduced to process 3D point clouds directly}. These models can generally be categorized into four types: point-wise MLP-based \cite{pointnet}\cite{pointnet++}\cite{pointmlp}, convolution-based\cite{pointcnn}\cite{pointconv}\cite{kpconv}, graph-based\cite{dgcnn}\cite{pointgnn}, and transformer-based models \cite{pct}\cite{pointtransformer}. 
An effective attack should be able to breach the broadest possible range of models.

To comprehensively assess various backdoor attacks targeting 3D point cloud networks, we select five widely used and illustrative victim models, including PointNet\cite{pointnet}, PointNet++\cite{pointnet++}, PointCNN\cite{pointcnn}, DGCNN\cite{dgcnn}, and PCT\cite{pct}, spanning the previously mentioned four categories. PointNet utilizes point-wise MLP and max-pooling to achieve an order-invariant global feature. PointNet++ improves \purple{upon} PointNet by incorporating multi-scale grouping to capture local details. PointCNN employs $\mathcal{X}$-conv operations to transform the point set into a potentially canonical order, making classical convolutional operations applicable. DGCNN constructs multi-scale graphs to extract representative features \purple{from} a 3D point cloud. PCT introduces an offset-attention mechanism along with an implicit Laplace operator and normalization refinement. This approach is inherently permutation-invariant, rendering it more suitable for point cloud learning than the original self-attention module in the Transformer framework\cite{transformer}.

\subsection{Backdoor Attack on Images}
Based on whether the trigger is uniform or conditioned on the input, backdoor attacks on images can be categorized into two main types: \textit{static} and \textit{sample-specific}. Representative static methods include BadNets\cite{badnets}, Blended\cite{blended}, and SIG\cite{sig}, while sample-specific methods encompass Input-Aware\cite{inputaware}, SSBA\cite{ssba}, etc. BadNet embeds a square patch on the image as the trigger. Blended replaces the trigger with a translucent watermark to reduce its conspicuousness. SIG contaminates the image in the frequency domain, and the pollution will naturally disperse throughout the image after being recovered to the spatial domain. Although static triggers are effective, their fixed patterns make them easy to identify. 

In contrast, sample-specific triggers, which depend on the input samples, are more diverse and thus harder to defend against. WaNet manually designs an invisible trigger by posing a tiny optical flow on the image. Input-Aware uses a network to generate an input-aware trigger. SSBA reconstructs the input image directly through the network. \bluesection{LIRA\cite{lira} and AdvDoor\cite{advdoor} take advantage of adversarial attack techniques for more powerful triggers.} These data-driven methods are more flexible and extensible.

\bluesection{Moreover, Reflection\cite{reflection} and PIB\cite{physical} consider natural triggers, such as light reflection and camera fingerprints, respectively, which are more friendly to physical deployment. }

\subsection{Backdoor Attack on 3D Point Clouds}
PointBA-I\cite{pointba} and PCBA (poisoned-label)\cite{pcba} \bluesection{simulate object interaction by attaching a cluster of points near the target object}. These \textit{interaction-based} methods often possess overwhelming attack performance, but the trigger's independence from the main body of the point cloud makes it easily filtered out by statistical methods.
\textit{Transformation-based} approaches view point clouds as geometric objects, employing rigid or non-rigid transformations as triggers. PointBA-O\cite{pointba} attacks point clouds through rotation, and NRBdoor\cite{nrbdoor} adapts this concept to 3D meshes. However, they can be nullified by rotation augmentation. IRBA\cite{irba} establishes \purple{several} local coordinate systems centered on the points with the farthest point sampling and performs scaling and rotation in each of them. Ultimately, these local affine transformations are compounded into a global one \purple{with} Gaussian smoothing. IRBA's trigger is nonlinear and sample-specific, making it hard to defend. However, similar to WaNet, IRBA's success also relies on ingenious manual design. Meanwhile, the irregularity of 3D point clouds increases the difficulty of manual trigger design. 

\begin{figure*}[!t]
\centering
\includegraphics[width=1.0\linewidth]{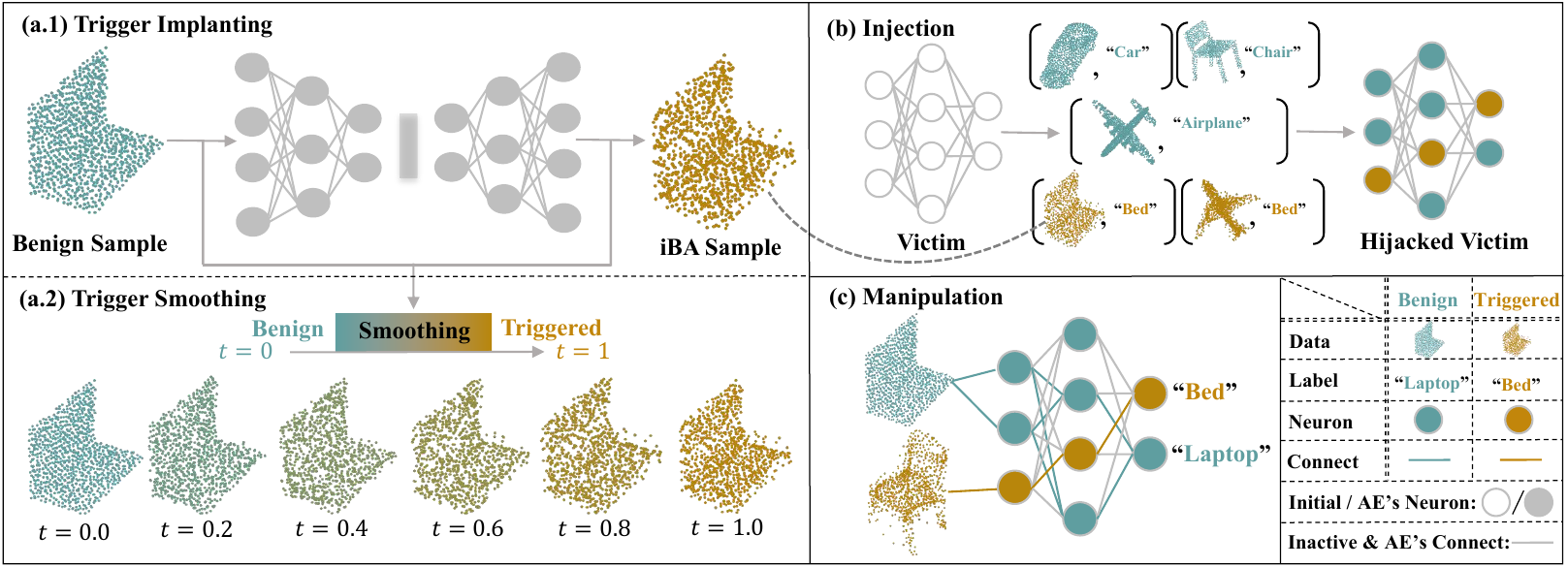}
\caption{The framework of our method. (a.1) Utilize a pre-trained AE to reconstruct benign samples and to generate backdoor samples; (a.2) An optional trigger smoothing module is designed to balance the imperceptibility and ASR. As $t$ shifts from 0 to 1, the polluted samples subtly morph from benign to backdoor ones; (b) The attacker manipulates the dataset by substituting a small portion of benign samples with backdoor ones and altering their labels to a specific target, such as "bed". The victim model is hijacked once trained on this manipulated dataset; (c) The hijacked model works normally under regular conditions but will misclassify a backdoor sample as a "bed" according to the attacker's predetermined plan. } 
\label{fig_1}
\end{figure*}
Differently, our iBA is proposed in the context of the successful experience in the image domain and the rapid development of point cloud reconstruction in recent years. It is a sample-specific and data-driven trigger with the advantages of effectiveness, imperceptibility, and \purple{defense penetration}. \purple{The attributes of different representative attacks are illustrated} in Tab.\ref{tab:types}. \bluesection{We skip PointBA-O when discussing sample-specific methods since the rotation is sample-agnostic, whereas the residual scene flow is not.} We also need to clarify the differences between our work and MorphNet\cite{morphnet}, as well as distinguish it from 3TPS\cite{3tps}. MorphNet obtains a covert trigger through an auxiliary network and iterative optimization. However, MorphNet focuses on \purple{a} clean-label task, which is not within our scope. 3TPS is based on trainable 3D Thin Plate Spline interpolation. \bluesection{It considers a white-box scenario where the attacker can modify the model or even the training process.}  
Additionally, Fan et al.\cite{mba} also consider 3D meshes and design a backdoor attack method based on remeshing.

\begin{figure}[!htbp]
\centering
\includegraphics[width=1.0\linewidth]{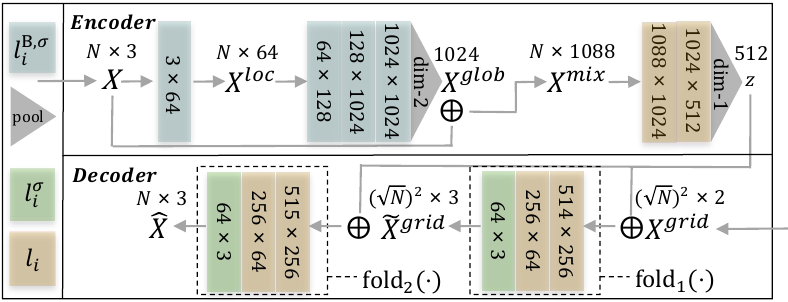}
\caption{Detailed structure of the utilized 3D AE. It is inspired by the design of FoldingNet\cite{foldingnet}. The blue, green, and yellow blocks correspond to three different types of neural network units, each annotated with the size of its parameter matrix $W$. The gray triangle represents a max-pooling layer, with dim-$i$ indicating the applied dimension. For definitions of the \purple{specific symbols}, please refer to Sec.\ref{mirror_1}.}
\label{ae}
\end{figure}

\begin{figure*}[!htbp]
\centering
\includegraphics[width=0.8\linewidth]{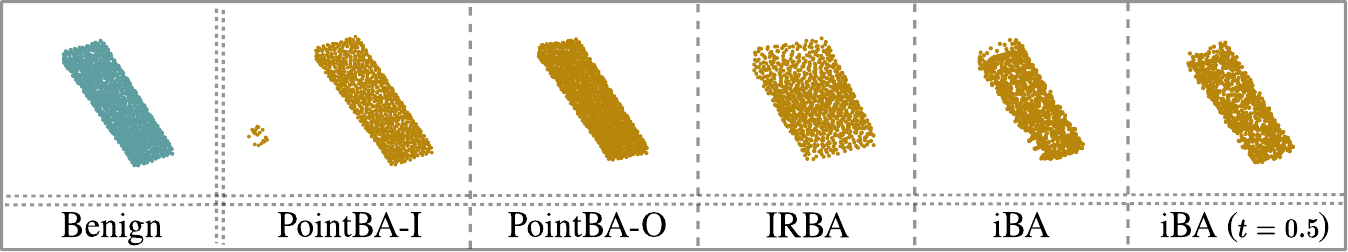}
\caption{Visualization of different attacks on a "keyboard" \purple{object} from the ModelNet40 dataset. It shows that the folding-based reconstruction introduces special geometric perturbations.} 
\label{method_display}
\end{figure*}

\section{Preliminaries}
\subsection{Threat Model}
The threat model delineates the capability boundary of the attacker and users. To simulate practical applications and make a fair comparison with previous methods, we consider \purple{a} classical scenario \purple{in} \cite{pointba} \cite{irba}, wherein users adopt a third-party dataset from an unknown source. In this context, the attacker contaminates a dataset and makes it available online. Then, the users train their models with this polluted dataset.
\bluesection{Finally, the attacker can create a backdoor entity through a 3D printer\cite{3dprint} and trigger the attack.}
Formally, the threat model can be defined as follows:

\begin{itemize}
\item{The attacker can manipulate the dataset, whereas they cannot modify the subsequent process, including the model structure, the training schedule, the deployment, and the inference.}
\item{\bluesection{The users' consciousness and knowledge about the attack can vary significantly. Some users might be completely oblivious to the attack, while others may realize its existence but lack exact information. A part of them can even partially or completely crack the attack. However, none of the users know the particular data being manipulated. We will clarify the users' abilities in the following discussions. }}
\end{itemize}

\subsection{Problem Formulation}

Consider a standard 3D point cloud classification task. Let $\mathcal{D}=\{(X_i, y_i)\}_{i=1}^{N}$ be a labeled dataset, \bluesection{where} $N$ is \bluesection{the} size of the dataset. Therein,  $X_i$ represents a point cloud with $n$ points and their Euclidean coordinates in $\mathbb{R}^3$ sampled from domain $\mathcal{X}$ lying in $\mathbb{R}^{n \times 3}$, i.e., $X_i \in \mathcal{X} \subset \mathbb{R}^{n \times 3}$. $y_i \in \mathcal{Y}=\purple{\{0, 1, \cdots, K-1\}}$ is the corresponding label and $K$ is the number of classes. $\mathcal{D}$ will be further divided into a training set $\mathcal{D}_{train}$ and a testing set $\mathcal{D}_{test}$. A point cloud classifier $f_{\boldsymbol{\theta}}: \mathcal{X} \rightarrow \mathcal{Y}$ parameterized with $\boldsymbol{\theta}$ will be trained on the $\mathcal{D}_{train}$.

The attacker then extracts a small portion of the data $\mathcal{D}_{b}$ in the training set and contaminates $\mathcal{D}_{train} = \mathcal{D}_{b} \cup (\mathcal{D}_{train} \backslash \mathcal{D}_{b})$. $\eta = |\mathcal{D}_{b}| / |\mathcal{D}_{train}|$ is the \textit{poisoning rate}, and $|\mathcal{D}_{b}|$ is the size of $\mathcal{D}_{b}$. Then, samples in $\mathcal{D}_{b}$ will be poisoned via a \textit{trigger implanting function} (TIF) $\mathcal{G}: \mathcal{X} \rightarrow \mathcal{X} \ $:

\begin{equation} \label{eq1}
    (X, y) \rightarrow (\mathcal{G}(X), y_t), \ \forall (X, y) \in \mathcal{D}_{b},
\end{equation}
where $y_t \in \mathcal{Y}$ is the target class. We denote the poisoned version of $\mathcal{D}_{b}$ as $\mathcal{D}'_{b}$, and the poisoned training set is \purple{denoted} as $\mathcal{D}'_{train}=\mathcal{D}'_{b} \cup (\mathcal{D}_{train} \backslash \mathcal{D}_{b})$. The attacker aims to hijack the classifier $f_{\boldsymbol{\theta}}$ when the users download $\mathcal{D}'_{train}$ and train $f_{\boldsymbol{\theta}}$ by minimizing the empirical risk:

\begin{equation} \label{eq2}
    \boldsymbol{\theta'}=\underset{\boldsymbol{\theta}} {\arg\min} \  \mathbb{E}_{(X,y) \sim \mathcal{P}_{\mathcal{D}'_{train}}} \left[\mathcal{L}\left(f_{\boldsymbol{\theta}}(X), y\right) \right],
\end{equation}
where $\mathcal{P}_{\mathcal{D}'_{train}}$ is the underlying distribution of $\mathcal{D}'_{train}$, and $\mathcal{L}: \mathcal{Y} \times \mathcal{Y} \rightarrow \mathbb{R}^{+}$ is the loss function. Ideally, the poisoned model $f_{\boldsymbol{\theta'}}$ will behave normally on benign samples to ensure the \textit{function preservation}:

\begin{equation} \label{acc}
    \mathbb{E}_{(X,y) \sim \mathcal{P}_{\mathcal{D}_{test}}} \left[\mathbb{I}\{f_{\boldsymbol{\theta}^{\prime}}(\boldsymbol{X}) = y\} \right],
\end{equation}
in which $\mathbb{I}\{ \cdot \}$ is the indicator function. Eq. \ref{acc} reflects the test accuracy of the poisoned model $f_{\boldsymbol{\theta'}}$. Higher test accuracy (ACC) indicates that the contaminated model is less affected on benign samples, reflecting the attack's stealthiness. On the other hand, \textit{attack success rate} (ASR) is defined as \purple{follows} to evaluate the effectiveness of the attack:
\begin{equation} \label{asr}
    \mathbb{E}_{(X,y) \sim \mathcal{P}_{\mathcal{D}_{test}}} \left[\mathbb{I}\{f_{\boldsymbol{\theta}^{\prime}}(\mathcal{A}(X)) = y_t\} \right],
\end{equation}
where $\mathcal{A}: \mathcal{X} \rightarrow \mathcal{X}$ is the \textit{trigger activation function}, which usually coincides with $\mathcal{G}$.

Moreover, we also emphasize the \textit{imperceptibility} of the trigger. Given a distance (or discrepancy) $d: \mathcal{X} \times \mathcal{X} \rightarrow [0,+\infty)$, consider:
\begin{equation} \label{imp}
    \mathbb{E}_{(X,y) \sim \mathcal{P}_{\mathcal{D}}} \left[d(\mathcal{G}(X), X) \right].
\end{equation}

\section{Our Method}
We introduce our method in two distinct parts: the trigger implanting function and the trigger smoothing module. Fig.\ref{fig_1} displays the overall framework of our method.

\subsection{iBA Trigger} \label{mirror_1}
The central challenge lies in devising a trigger implanting function $\mathcal{G}$ for 3D point clouds. $\mathcal{G}$ must produce backdoor samples that are not only imperceptible but also \purple{effective}.
In this work, we denote the trigger implanting function of our iBA as $\mathcal{G}_{iBA}$. We formulate $\mathcal{G}_{iBA}$ with a folding-based AE\cite{foldingnet}. 
In detail, given an input 3D point cloud $X \in \mathcal{X}  \subset \mathbb{R}^{1024 \times 3}$, we first embed $X$ to a compact latent code $z \in \mathcal{Z} \subset \mathbb{R}^{512}$ with an encoder $\psi: \mathcal{X} \rightarrow \mathcal{Z}$:
\begin{equation} \label{enc}
    z = \psi(X).
\end{equation}
For the sake of brevity, we \purple{denote} the three different types of layers utilized in the network as:
\begin{gather} \label{layername}
l_{i}^{\mathrm{B}, \mathrm{\sigma}}(X)=\operatorname{ReLU}(\operatorname{BN}(W_{i}^{\mathrm{B}, \mathrm{\sigma}}X+b_{i}^{\mathrm{B}, \mathrm{\sigma}})), \\
l_{i}^{\mathrm{\sigma}}(X)=\operatorname{ReLU}(W_{i}^{\mathrm{\sigma}}X+b_{i}^{\mathrm{\sigma}}), \\
l_{i}(X)=W_{i}X+b_{i},
\end{gather}
\bluesection{where $\operatorname{BN}(\cdot)$ and $\operatorname{ReLU}(\cdot)$ are batch normalization (BN)\cite{batchnorm} and rectified linear unit (ReLU)\cite{relu}, abbreviated as $B$ and $\sigma$ respectively in the superscripts of network parameters $W$ and $b$.}

We \bluesection{construct} the encoder $\psi(\cdot)$ using a five-layer perceptron architecture, where three layers are dedicated to extracting point-wise features $X^{loc}$ and global features $X^{glob}$ from the input point cloud:
\begin{gather} \label{encoder-step1}
X^{loc} = l_{1}^{\mathrm{B}, \mathrm{\sigma}}(X), \\
X^{glob} = \operatorname{MaxPool}(l_{3}^{\mathrm{B}, \mathrm{\sigma}}(l_{2}^{\mathrm{B}, \mathrm{\sigma}}(X^{loc}))), 
\end{gather}
with $\operatorname{MaxPool}(\cdot)$ representing the max-pooling operation. We then concatenate $X^{loc}$ with $X^{glob}$ for each point, forming a hybrid feature $X^{mix}$  that captures both local and global contexts. Subsequently, we use two additional layers to derive the latent representation $z$:
\begin{gather} \label{encoder-step2}
X^{mix} = [X^{loc};X^{glob}], \\
z = \operatorname{MaxPool}(l_{2}^{\mathrm{\sigma}}(l_{1}^{\mathrm{\sigma}}(X^{mix}))).
\end{gather}

For the decoder, we initialize its output as a 2D grid $X^{grid} \in \mathbb{R}^{32 \times 32 \times 2}$, centered at the origin. This grid, when unfolded along its third channel, forms a 3D point cloud $X^{grid} \in \mathcal{X} \subset \mathbb{R}^{1024 \times 3}$ positioned on the $z=0$ plane. The encoded geometric details in $z$ are incrementally infused into the grid $X^{grid}$ through a folding mechanism $\operatorname{fold}: \mathcal{X} \times \mathcal{Z} \rightarrow \mathcal{X}$. This folding operation is performed twice to ensure that $X^{grid}$ fully assimilates the global feature $z$ of the point cloud, \purple{thereby} achieving \purple{improved reconstruction quality}.

\begin{equation} \label{foldingnet}
    \hat{X} = \operatorname{fold}_2(\operatorname{fold}_1(X^{grid},z),z),
\end{equation}
where $\operatorname{fold}_j, \;j=1,2$ are defined as:
\begin{equation} \label{fold}
    \operatorname{fold}_j(X,z) = l_{4,j}^{\mathrm{\sigma}}(l_{3,j}^{\mathrm{\sigma}}([X;z])).
\end{equation}
Please refer to Fig.\ref{ae} for the detailed structure of the entire AE.

Finally, we have our trigger implanting function $\mathcal{G}_{\purple{iBA}}: \mathcal{X} \rightarrow \mathcal{X}$:
\begin{equation} \label{mirror}
    \mathcal{G}_{\purple{iBA}}(X) = \operatorname{fold}_2^{*}(\operatorname{fold}_1^{*}(X^{grid},\psi^{*}(X)),\psi^{*}(X)),
\end{equation}
where $\operatorname{fold}_j^{*}(\cdot,\cdot)$ and $\psi^{*}(\cdot)$ denote the well-trained folding modules and encoder respectively. 

The reconstruction loss of the AE guides the appearance of the backdoor shape and, in turn, impacts the imperceptibility and effectiveness.
We additionally introduce the Wasserstein distance\cite{wd} aside from the standard Chamfer discrepancy\cite{cd} $d_{CD}: \mathcal{X} \times \mathcal{X} \rightarrow [0, +\infty)$ for training the AE:
\begin{equation} \label{loss}
    \mathcal{L}_{AE} = \mathbb{E}_{X \sim \mathcal{D}}[ \lambda_{1}d_{CD}(X, \hat{X}) + \lambda_{2}d_{SWD}(X, \hat{X}) ],
\end{equation}
where $d_{SWD}: \mathcal{X} \times \mathcal{X} \rightarrow  [0, +\infty)$ is the sliced version of Wasserstein distance\cite{swd} based on Radon Transform for accelerating and $\lambda_{1}, \lambda_{2} > 0$ are balancing weights. 
The involvement of $d_{SWD}$ constrains the overall point distribution, which not only further improves the imperceptibility but also optimizes the geometric perturbations in the backdoor shape, resulting in \purple{enhanced effectiveness against} PointNet. Please refer to Sec.\ref{AE_achitecture} and Sec.\ref{pattern} for more discussion on the choice of $\mathcal{L}_{AE}$ and network architecture.
 
\subsection{Trigger Smoothing}

\begin{figure*}[!htbp]
\centering
\includegraphics[width=0.9\linewidth]{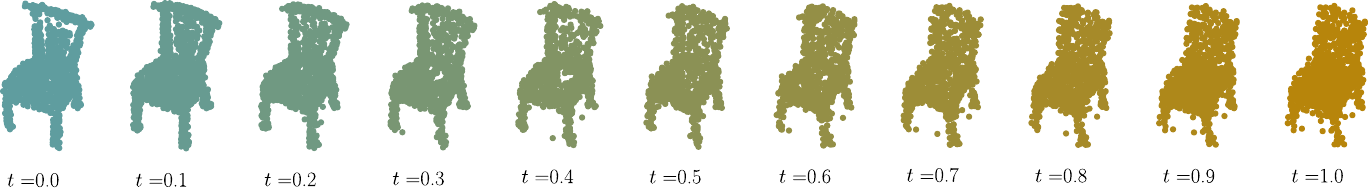}
\caption{Visualization of trigger smoothing. As $t$ increases from 0 to 1, the benign point cloud is gradually transformed into a polluted one.}
\label{smooth}
\end{figure*}

\begin{figure}[!htbp]
\centering
\includegraphics[width=1.0\linewidth]{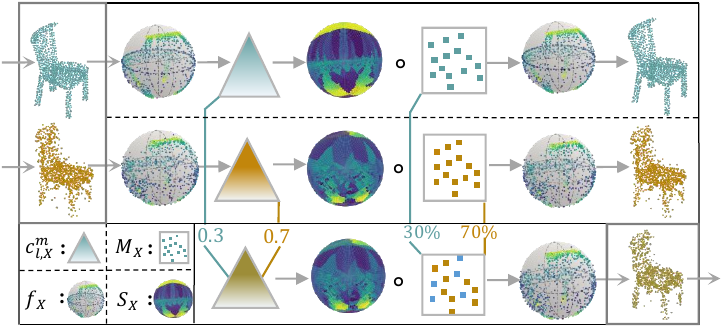}
\caption{The process of trigger smoothing. The benign and triggered point clouds are initially transformed into spherical functions. Then, discrete SHT derives \purple{the} harmonic coefficients, which facilitate the reconstruction of underlying surfaces via the inverse transformation. Sampling the underlying surface with the location matrix will reproduce the point cloud. Furthermore, the linear interpolation among these harmonic coefficients allows for a homotopy of the underlying surfaces. Finally, sampling the interpolated underlying surface with a randomly mixed location matrix generates the interpolated point cloud.} 
\label{smoothpipeline}
\end{figure}

The iBA trigger has a fixed attack intensity hinged on the reconstruction error from the trained AE $\mathcal{G}_{iBA}$. Thus, we should introduce a trigger smoothing module to provide controllable triggers. Given a TIF $\mathcal{G}$, the core of the trigger smoothing problem is to establish a \textit{homotopy} $H_\mathcal{G}$ from the benign point cloud $X$ to the triggered one $\mathcal{G}(X)$:
\begin{gather} \label{triggersmooth}
H_\mathcal{G}(\cdot\;;X): [0,1] \rightarrow \mathcal{X}, \\
s.t. \; H_\mathcal{G}(0;X)=X,\; H_\mathcal{G}(1;X)=\mathcal{G}(X).
\end{gather}
$H_{\mathcal{G}_{iBA}}$ is nontrivial since the folding-based AE lacks an explicit trace of deformation. We propose to use spherical harmonic transformation (SHT) to forge the connection between the $X$ and $\mathcal{G}_{iBA}$ in the frequency domain. A 3D point cloud $X$ can be interpreted as a discrete spherical function $f_X:\mathbb{S}^2 \rightarrow \mathbb{R}$. For a specific longitude $\phi \in (0, 2\pi]$ and latitude $\theta \in [0, \pi]$, if a point exists in direction $(\phi, \theta)$, $f_X(\phi, \theta)$ is the directed projected length from that point to the unit sphere (take the \purple{nearest one} if multiple points exist), otherwise, $f_X(\phi, \theta)=0$. We apply the discrete spherical harmonic transform (DSHT) on the discrete spherical function $f_X$:
\begin{gather} \label{DSHT}
c_{l,X}^m = f_X(\phi, \theta) \cdot  (Y_l^m(\phi, \theta) \circ M_X), \\
Y_l^m(\phi, \theta)=\sqrt{\frac{(2 l+1)}{4 \pi} \frac{(l-m) !}{(l+m) !}} P_l^m(\cos \theta) e^{i m \phi}.
\end{gather}
where $c_{l,X}^m$ and $P_l^m(\cdot)$ are the spherical harmonic coefficients and the associated Legendre polynomials. $\cdot$ and $\circ$ denote the inner and Hadamard products, respectively. $M_X$ is a sparse binary mask indicating the non-zero elements of $f_X(\phi, \theta)$, namely the point locations on the celestial sphere. Then, we further apply the inversed spherical harmonic transform (ISHT) on $\{c_{l,X}^m\}$ to reconstruct the underlying surface of $X$:
\begin{equation} \label{surface}
    S_X = \sum_{l=0}^{N_l} \sum_{m=-l}^l c_{l,X}^m Y_l^m(\phi, \theta),
\end{equation}
where $N_l$ is the highest order of the transformation and $S_X$ is an approximation of the underlying surface. Naturally, we can reproduce $f_X$ by:
\begin{equation} \label{reproduce}
    f_X \approx S_X \circ M_X,
\end{equation}
Eq.\eqref{reproduce} decomposes the point cloud into a discrete location mask $M_X$, and a \purple{continuous} and deformable underlying surface $S_X$. By interpolating the coefficients, we can establish a trivial homotopy on the latter, simplifying the connection between $X$ and $\mathcal{G}(X)$:

\begin{gather} \label{homotopy}
c_{l,t}^m=(1-t)c_{l,X}^m+tc_{l,\mathcal{G}(X)}^m, \\
M_t = \operatorname{sp}(M_X;1-t)+\operatorname{sp}(M_{\mathcal{G}(X)};t),
\end{gather}
where $\operatorname{sp}(M_X;1-t)$ represents uniformly sampling the non-zero elements of matrix $M_X$ in proportion to $1-t$. Finally, we have $H_{\mathcal{G}}$:
\begin{gather} \label{hg_final}
    H_\mathcal{G}(t;X)=S_t \circ M_t, \\
    S_t = \sum_{l=0}^{N_l} \sum_{m=-l}^l c_{l,t}^m Y_l^m(\phi, \theta).
\end{gather}
Fig.\ref{smoothpipeline} briefly describes the process of trigger smoothing, and Fig.\ref{smooth} shows an intuitive example of trigger smoothing.

It is noteworthy that, compared to directly mixing the point sets, establishing a homotopy on the underlying surface allows for smooth deformation of the point cloud shape. However, the difficulty is further shifted to the correspondence of location masks. The rough blending of location masks might still result in changes in the local point cloud distribution and cause unevenness. \purple{This limitation is manageable when $\mathcal{G}=\mathcal{G}_{iBA}$ since the AE can reconstruct the target point cloud without significant posture change. }

\section{Experiments} \label{experiments}

\subsection{Experiment Setup}
\subsubsection{Datasets and Victim Models}
The proposed method is evaluated on 3D shape classification tasks. In our experiments, we use three commonly used benchmarks, i.e., ModelNet10, ModelNet40, and ShapeNetPart. For ModelNet40, we adopt the official split of 9843 point clouds for training and 2468 for testing, respectively. ModelNet10 is a reduced version of ModelNet40 containing 10 categories. ShapeNetPart is a portion of ShapeNet \bluesection{with} 12128 training shapes and 2874 testing shapes with 16 categories in total. We uniformly sample 1024 points from each shape and normalize the point cloud into $[-1,1]^3$. \purple{As} victim classifiers, we inherit the most widely used PointNet\cite{pointnet}, PointNet++\cite{pointnet++}, PointCNN\cite{pointcnn}, and DGCNN\cite{dgcnn} in previous work. Considering the rapid development of transformer-based point cloud classifiers\cite{pct}\cite{pointtransformer} recently, we additionally introduce PCT\cite{pct} as our new victim. 

\begin{table*}
\begin{center}
\caption{Comparison of our iBA with other methods in terms of ACC and ASR on five victim models (PointNet, PointNet++, DGCNN, PCT, and PointCNN) across three different datasets (ModelNet40, ModelNet10, and ShapeNetPart). Our method achieves outstanding ASRs } 
\label{tab:asr}
\renewcommand \arraystretch{1.2}
\resizebox{0.9\linewidth}{!}{
\begin{tabular}{c|c|cc|cc|cc|cc|cc}
\hline
\multirow{2}{*}{Datasets} & \multirow{2}{*}{Methods} & \multicolumn{2}{c|}{PointNet}          & \multicolumn{2}{c|}{PointNet++}        & \multicolumn{2}{c|}{DGCNN}       & \multicolumn{2}{c|}{PCT}        & \multicolumn{2}{c}{PointCNN}      \\ \cline{3-12} 
                         &                         & \multicolumn{1}{c|}{ACC$\uparrow$}  & ASR$\uparrow$  & \multicolumn{1}{c|}{ACC$\uparrow$}  & ASR$\uparrow$  & \multicolumn{1}{c|}{ACC$\uparrow$}  & ASR$\uparrow$  & \multicolumn{1}{c|}{ACC$\uparrow$} & ASR$\uparrow$ & \multicolumn{1}{c|}{ACC$\uparrow$} & ASR$\uparrow$ \\ \hline \hline
\multirow{4}{*}{ModelNet40}     & PointBA-I                   & \multicolumn{1}{c|}{89.6} & 99.8 & \multicolumn{1}{c|}{91.3} & 100 & \multicolumn{1}{c|}{90.6}  & 100 & \multicolumn{1}{c|}{90.6} & 100 & \multicolumn{1}{c|}{91.2} & 100                              \\ \cline{2-12} 
                         & PointBA-O                   & \multicolumn{1}{c|}{88.7} & 78.2 & \multicolumn{1}{c|}{91.0} & 91.3 & \multicolumn{1}{c|}{91.0} & 82.0 & \multicolumn{1}{c|}{90.5}& 81.7 & \multicolumn{1}{c|}{91.2} & 82.9 \\ \cline{2-12} 
                         & IRBA                    & \multicolumn{1}{c|}{88.2} & 81.3 & \multicolumn{1}{c|}{91.4} & 95.4 & \multicolumn{1}{c|}{90.6} & 81.9 & \multicolumn{1}{c|}{89.5}& 74.2 & \multicolumn{1}{c|}{91.3} & 77.0 \\ \cline{2-12} 
                         & iBA(Ours)            & \multicolumn{1}{c|}{89.4} & 82.0 & \multicolumn{1}{c|}{91.2} & 99.7 & \multicolumn{1}{c|}{91.0} & 99.6 & \multicolumn{1}{c|}{90.5}& 98.8& \multicolumn{1}{c|}{91.6} & 99.8 \\ \hline \hline
\multirow{4}{*}{ModelNet10}     & PointBA-I                   & \multicolumn{1}{c|}{92.5} & 100  & \multicolumn{1}{c|}{92.6} & 100  & \multicolumn{1}{c|}{93.5} & 100  & \multicolumn{1}{c|}{93.4}& 100 & \multicolumn{1}{c|}{93.3} &  100\\ \cline{2-12} 
                         & PointBA-O                   & \multicolumn{1}{c|}{91.7} & 79.6 & \multicolumn{1}{c|}{92.6} & 95.0 & \multicolumn{1}{c|}{93.1} & 86.0 & \multicolumn{1}{c|}{93.6}& 86.6 & \multicolumn{1}{c|}{93.5} & 84.3 \\ \cline{2-12} 
                         & IRBA                    & \multicolumn{1}{c|}{92.9} & 83.1 & \multicolumn{1}{c|}{93.5} & 92.8 & \multicolumn{1}{c|}{92.4} & 81.6 & \multicolumn{1}{c|}{92.8}& 75.9 & \multicolumn{1}{c|}{92.3} & 76.9 \\ \cline{2-12} 
                         & iBA(Ours)            & \multicolumn{1}{c|}{91.7} & 81.3 & \multicolumn{1}{c|}{93.1} & 100  & \multicolumn{1}{c|}{94.0} & 99.5 & \multicolumn{1}{c|}{93.8}& 99.1 & \multicolumn{1}{c|}{93.4} & 100 \\ \hline \hline
\multirow{4}{*}{ShapeNetPart}   & PointBA-I                   & \multicolumn{1}{c|}{98.5} & 100  & \multicolumn{1}{c|}{98.9} & 100  & \multicolumn{1}{c|}{98.8} & 100  & \multicolumn{1}{c|}{98.7} & 100 & \multicolumn{1}{c|}{98.3} & 100 \\ \cline{2-12} 
                         & PointBA-O                   & \multicolumn{1}{c|}{98.3} & 92.2 & \multicolumn{1}{c|}{98.8} & 92.6 & \multicolumn{1}{c|}{98.9} & 88.9 & \multicolumn{1}{c|}{98.8} & 83.5& \multicolumn{1}{c|}{98.3} & 86.6  \\ \cline{2-12} 
                         & IRBA                    & \multicolumn{1}{c|}{98.2} & 91.3 & \multicolumn{1}{c|}{98.9} & 99.9 & \multicolumn{1}{c|}{98.7} & 83.0 & \multicolumn{1}{c|}{98.1} & 84.6 & \multicolumn{1}{c|}{97.6} & 88.0  \\ \cline{2-12} 
                         & iBA(Ours)            & \multicolumn{1}{c|}{98.3} & 94.6 & \multicolumn{1}{c|}{98.9} & 99.7 & \multicolumn{1}{c|}{98.9} & 99.6 & \multicolumn{1}{c|}{98.8} & 99.7 & \multicolumn{1}{c|}{98.3} & 100  \\ \hline
\end{tabular}
}
\end{center}
\end{table*}

\begin{table}
\begin{center}
\caption{Baseline accuracy of the reproduced victim models. All attack methods in Tab.\ref{tab:asr}, including our iBA, do not significantly sacrifice the benign accuracy}
\label{tab:acc}
\renewcommand \arraystretch{1.2}
\resizebox{0.9\linewidth}{!}{
\begin{tabular}{c|c|c|c}
\hline
Models & ModelNet40 & ModelNet10 & ShapeNetPart \\ \hline \hline
PointNet     & 88.9       & 92.8       & 98.5        \\ \hline
PointNet++   & 91.4       & 93.7       & 99.0         \\ \hline
DGCNN  & 91.1       & 94.0       & 98.8         \\ \hline
PCT    & 90.8       & 93.9       & 98.9         \\ \hline
PointCNN   & 91.4       & 93.3       & 98.4         \\ \hline
\end{tabular}
}
\end{center}
\end{table}

\subsubsection{Trigger Generation}

We compare our iBA method with three representative 3D point cloud backdoor attacks: PointBA-I\cite{pointba}, PointBA-O\cite{pointba}, and IRBA\cite{irba}. 
We reproduce PointBA-I by inserting a sphere-like cluster with a fixed radius of 0.1, positioned at $(-0.9, -0.9, -0.9)$ on the normalized axis, following the specifications detailed in \cite{pointba}. The cluster comprises 2\% of the total points. For PointBA-O, we configure the Euler rotation angles to $(0^{\circ}, 0^{\circ}, 10^{\circ})$. For IRBA, we use 16 seed points, localized rotations of $5^{\circ}$, and scaling with factor 5 before the point cloud is normalized.
In our iBA, we adjust the dimension of $z$ to 512 and set $\lambda_1=1.0, \lambda_2=0.001$. The model is trained with Adam optimizer\cite{adam} for 300 epochs, with a learning rate of 0.001, batch size \purple{of} 16.
Theoretically, when $N_l$ is large enough, the transformation error can be adequately small. However, $N_l=100$ has been satisfactory in practice. For the endpoint cases $t=0$ and $t=1$, we directly set $H_{\mathcal{G}_{\purple{iBA}}}(0;X)=X$ and $H_{\mathcal{G}_{\purple{iBA}}}(1;X)=\mathcal{G}_{\purple{iBA}}(X)$.

\subsubsection{Trigger Implanting}
The poisoning rate $\eta$ is 0.02 in our experiments. For ModelNet10, ModelNet40, and ShapeNetPart, the target labels are "Table" ("8"), "Toilet" ("35"), and "Lamp" ("8"), respectively, \purple{following} our pioneers\cite{pointba}\cite{irba}. Backdoor candidates are randomly selected from different classes. Victim models are trained on the backdoor dataset using the Adam optimizer for 200 epochs, with a learning rate of 0.001 and batch size of 16.
\subsubsection{Evaluation Metrics}
We adopt ASR to evaluate the effectiveness of backdoor attacks. In addition, stealthiness is also an important indicator, which consists of two aspects: functional preservation and imperceptibility. Functional preservation focuses on whether the backdoor model remains accurate on benign samples. Imperceptibility concerns the level of deviation between the source samples and the backdoor ones. Previous work generally utilizes Chamfer discrepancy (CD)\cite{cd} to depict imperceptibility. We enrich the metric\purple{s} with Wasserstein distance (WD, or Earth Mover's distance, EMD)\cite{wd} and Hausdorff distance (HD)\cite{hd} for a more comprehensive evaluation. 

All our experiments are conducted on a single GeForce RTX 2080Ti GPU.

\begin{table}
\begin{center}
\caption{Quantitative analysis reveals the imperceptibility of our iBA through mean errors in CD, WD, and HD on three datasets} 
\label{tab:distance}
\renewcommand \arraystretch{1.3}
\resizebox{1.0\linewidth}{!}{
\begin{tabular}{c|c|c|c|c}
\hline
\multicolumn{1}{l|}{Datasets}  & Methods       & CD  $\times 100$$\downarrow$ & WD $\times 0.1$$\downarrow$ & $\;$ HD$\downarrow$ $\;$ \\ \hline \hline
\multirow{4}{*}{ModelNet40}          & PointBA-I        &  0.41               & 1.81    & 0.47     \\ \cline{2-5} 
                              & PointBA-O        &  0.13               & 0.61    & 0.07     \\ \cline{2-5} 
                              & IRBA         &  0.47               & 2.05    & 0.14     \\ \cline{2-5} 
                              & iBA (Ours) &  0.20               & 0.83    & 0.13     \\ \hline \hline
\multirow{4}{*}{ModelNet10}          & PointBA-I        &  0.33               & 1.83    & 0.43    \\ \cline{2-5} 
                              & PointBA-O        &  0.12               & 0.50    & 0.06     \\ \cline{2-5} 
                              & IRBA         &  0.43               & 2.05    & 0.14     \\ \cline{2-5} 
                              & iBA (Ours) &  0.20               & 0.81    & 0.14     \\ \hline \hline
\multirow{4}{*}{ShapeNetPart}        & PointBA-I        &  0.45               & 1.72    & 0.50    \\ \cline{2-5} 
                              & PointBA-O        &  0.14               & 0.58    & 0.07    \\ \cline{2-5} 
                              & IRBA         &  0.41               & 1.79    & 0.14   \\ \cline{2-5} 
                              & iBA (Ours) &  0.13               & 0.69    & 0.13  \\ \hline
\end{tabular}
}
\end{center}
\end{table}

\begin{table*}
\begin{center}
\caption{The penetrative power of different methods against seven types of data augmentation and two special defense techniques applied to DGCNN on ModelNet10 is evaluated. Both PointBA-I and PointBA-O have critical weaknesses, while IRBA and our iBA can effectively penetrate various defenses} 

\label{tab:def}
\renewcommand \arraystretch{1.2}
\resizebox{0.9\linewidth}{!}{
\begin{tabular}{c|c|cc|cc|cc|cc}
\hline
\multirow{2}{*}{Type}         & \multirow{2}{*}{Methods} & \multicolumn{2}{c|}{PointBA-I} & \multicolumn{2}{c|}{PointBA-O} & \multicolumn{2}{c|}{IRBA}      & \multicolumn{2}{c}{iBA(Ours)} \\ \cline{3-10} 
                              &                          & \multicolumn{1}{c|}{ACC$\uparrow$} & ASR$\uparrow$ & \multicolumn{1}{c|}{ACC$\uparrow$} & ASR$\uparrow$ & \multicolumn{1}{c|}{ACC$\uparrow$} & ASR$\uparrow$ & \multicolumn{1}{c|}{ACC$\uparrow$}   & ASR$\uparrow$  \\ \hline \hline
\multirow{7}{*}{Augmentation} & R                        & \multicolumn{1}{c|}{93.1} & 99.9 & \multicolumn{1}{c|}{92.6} & \textbf{0.0}  & \multicolumn{1}{c|}{93.6} & 78.7 & \multicolumn{1}{c|}{93.0} & 100  \\ \cline{2-10} 
                              & R3                       & \multicolumn{1}{c|}{93.6} & 100  & \multicolumn{1}{c|}{93.0} & \textbf{0.0}  & \multicolumn{1}{c|}{92.9} & 53.8 & \multicolumn{1}{c|}{93.9} & 100  \\ \cline{2-10} 
                              & Scaling                  & \multicolumn{1}{c|}{94.1} & 100  & \multicolumn{1}{c|}{93.5} & 86.0 & \multicolumn{1}{c|}{94.0} & 80.4 & \multicolumn{1}{c|}{93.9} & 99.8  \\ \cline{2-10} 
                              & Translation              & \multicolumn{1}{c|}{93.1} & 99.6 & \multicolumn{1}{c|}{92.4} & 85.5 & \multicolumn{1}{c|}{92.3} & 77.8 & \multicolumn{1}{c|}{92.9} & 100 \\ \cline{2-10} 
                              & Dropout                  & \multicolumn{1}{c|}{93.3} & 98.5 & \multicolumn{1}{c|}{93.5} & 90.6 & \multicolumn{1}{c|}{93.8} & 85.9 & \multicolumn{1}{c|}{93.2} & 99.3 \\ \cline{2-10} 
                              & Jitter                   & \multicolumn{1}{c|}{93.0} & 100  & \multicolumn{1}{c|}{93.1} & 83.2 & \multicolumn{1}{c|}{92.4} & 83.0 & \multicolumn{1}{c|}{93.2} & 99.6  \\ \cline{2-10} 
                              & SOR                      & \multicolumn{1}{c|}{92.6} & \textbf{0.0}  & \multicolumn{1}{c|}{93.1} & 86.3 & \multicolumn{1}{c|}{92.0} & 81.6 & \multicolumn{1}{c|}{92.7} & 98.9  \\ \hline \hline
\multirow{2}{*}{Defense}      & DUP                      & \multicolumn{1}{c|}{92.7} & 100  & \multicolumn{1}{c|}{92.6} & 85.2 & \multicolumn{1}{c|}{87.8} & 91.1 & \multicolumn{1}{c|}{89.7} & 89.7 \\ \cline{2-10} 
                              & LPF                      & \multicolumn{1}{c|}{92.6} & 99.6 & \multicolumn{1}{c|}{92.5} & 80.4 & \multicolumn{1}{c|}{93.0} & 77.1 & \multicolumn{1}{c|}{93.0} & 88.3 \\ \hline

\end{tabular}
}
\end{center}
\end{table*}

\subsection{Performance of Our Method}
We comprehensively compare the performance of PointBA-I, PointBA-O, IRBA, and our iBA in effectiveness and imperceptibility against multiple types of victim models on three different datasets.

\subsubsection{Effectiveness and Functional Preservation Analysis} 

Tab.\ref{tab:asr} presents the ASR and the clean test accuracy (ACC) of the victim models concerning different backdoor attacks. Experimental results verify that our iBA achieves a marvelous ASR across all cases. Although our ASR is slightly lower than PointBA-I, we have overcome some shortcomings of PointBA-I like low imperceptibility and high vulnerability under defense measures, which will be further analyzed in \ref{resist}.  
We also observe that, compared to PointNet, our iBA demonstrates superior performance on advanced models such as PointNet++, DGCNN, PointCNN, and PCT. This suggests that these models, which place a greater emphasis on the geometric details within the point cloud, are more sensitive to iBA.

\begin{table}
\begin{center}
\caption{PointCRT's recognition accuracy on different attacks in ModelNet10. Our iBA exhibits a comparable low exposure rate} 
\label{pointcrt}
\renewcommand \arraystretch{1.2}
\resizebox{0.9\linewidth}{!}{
\begin{tabular}{c|c|c|c|c}
\hline
        & PointBA-I  & PointBA-O  & IRBA & iBA(Ours)  \\ \hline \hline
F1$\downarrow$      & 0.972       & 0.721       & 0.899 & 0.804          \\ \hline
AUC$\downarrow$     & 0.997       & 0.760       & 0.960 & 0.877          \\ \hline

\end{tabular}
}
\end{center}
\end{table}

Tab.\ref{tab:acc} presents the baseline accuracy. Empirically, the compromised dataset is a detrimental factor\cite{statistical}. However, most backdoor models, including iBA, \purple{can approach the} baseline accuracy, with a few even slightly exceeding it. We conjecture that the low poisoning rate significantly mitigates the performance degradation induced by \purple{the} compromised data. Consequently, the influence of dataset disturbance \purple{may} occasionally outweigh the remaining side effects.
To wit, the iBA backdoor models remain discriminative on benign samples, verifying the stealthiness. 

\subsubsection{Imperceptibility Analysis} 
If imperceptibility is not limited, pursuing a high ASR becomes a competition for destruction rather than an art of balance. Therefore, imperceptibility is another important dimension in measuring a trigger. Tab.\ref{tab:distance} \bluesection{quantitatively} measures the intensity of different backdoor triggers by computing the average deviations under multiple metrics to comprehensively evaluate the imperceptibility. CD is sensitive to outliers but robust against small perturbations. HD reflects the worst-case scenario, and WD is suitable for comparing two point clouds that might differ in density. 
Compared to \purple{another robust attack} IRBA, we have significantly improved the imperceptibility under CD and WD. Among all methods, we are only second to the rotation-based PointBA-O. This indicates that iBA is not only effective and robust but also imperceptible.

\begin{figure}[!htbp]
\centering
\includegraphics[width=0.9\linewidth]{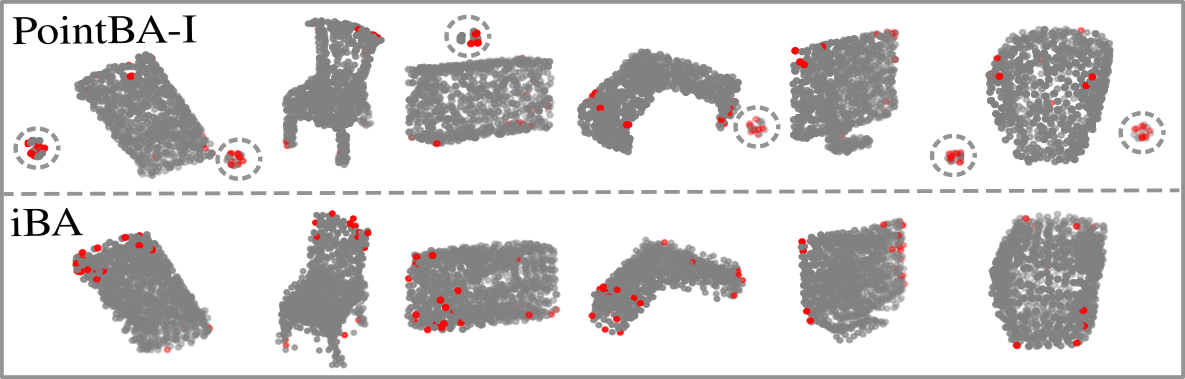} 
\caption{Gradient-based salience analysis with \cite{salience}. Red points are highly significant. The ball patterns (marked with dotted circles) \purple{of} PointBA-I are easily identifiable, whereas our iBA is challenging to locate due to its global nature.} 
\label{saliency}
\end{figure}

\subsection{Resistance to Defense Measures} \label{resist}
In practical applications, data augmentation is a standard technique to improve model robustness. A backdoor attack is invasive enough only if it can penetrate commonly used data augmentation methods. Moreover, some defense techniques\cite{dup}\cite{lpf}\cite{pointcrt} are specifically designed to mitigate the invasion of attacks, posing new requirements for the robustness of attacks. Unless otherwise mentioned, the experimental subjects in this subsection are DGCNN and ModelNet10.
\subsubsection{Resistance to Data Augmentation}

\begin{figure*}[!htbp]
\centering
\includegraphics[width=1.0\linewidth]{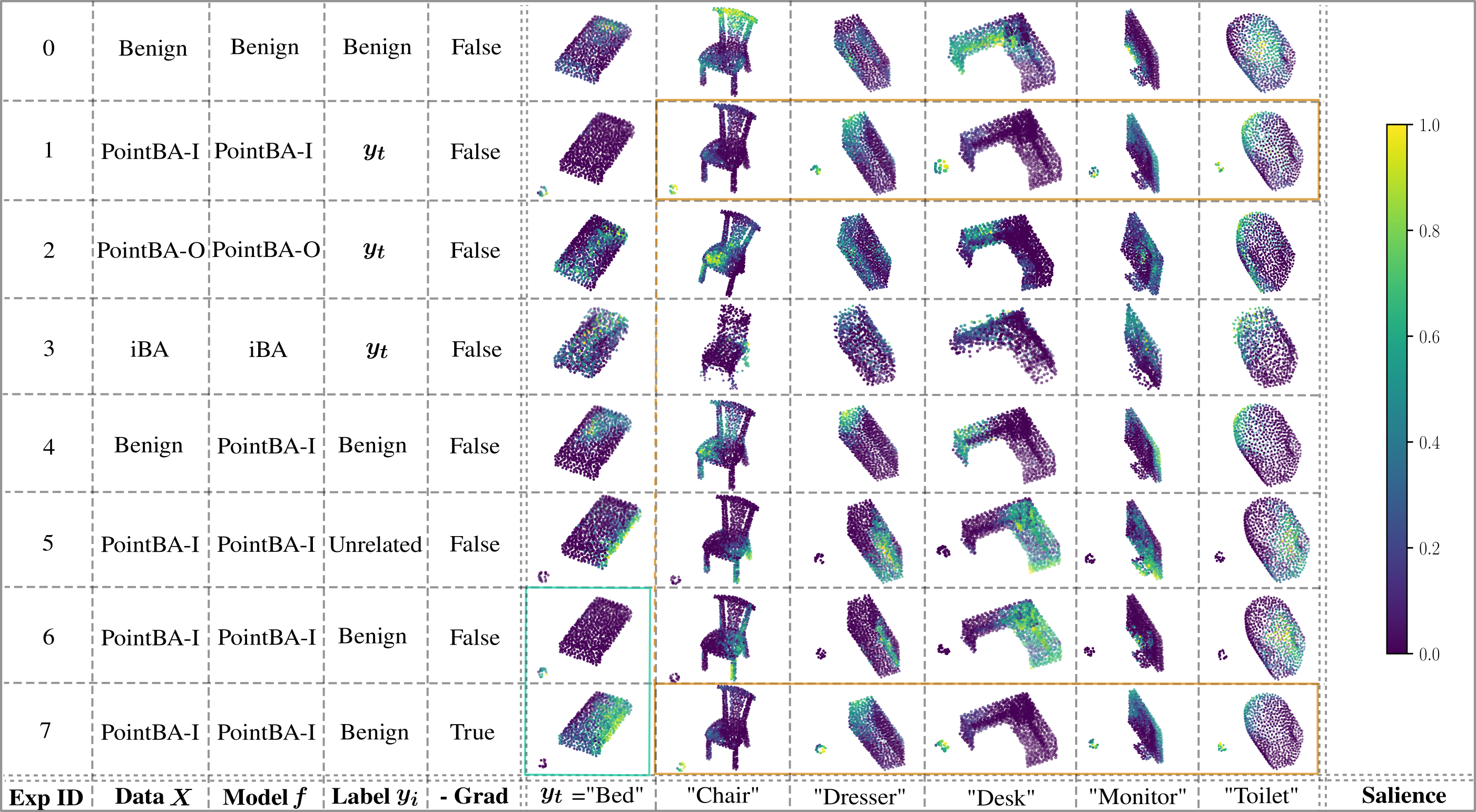}
\caption{3D Grad-CAM for the cross-display of different point clouds $X$, models $f$, and reference labels $y_i$. Model "PointBA-I/O" indicates that it has been hijacked with the PointBA-I/O attack. "-Grad" \purple{specifies} whether (True) or not (False) we inverse the gradient for a counterfactual study.
Exp.0-3 shows the difficulty of locating the \bluesection{dispersed} patterns. Exp.4 demonstrates that the hijack diminishes the model's ability to display salient areas of benign data. The blue box reveals a competitive relationship between the trigger and the source feature. The high consistency between the two yellow boxes mutually validates this competitive relationship. For a detailed analysis, please refer to Sec.\ref{cam-section}.}
\label{grad-cam}
\end{figure*}

The data augmentation techniques evaluated in our study include Rotation (R), Rotation-3D (R3), Scaling, Translation, Dropout, Jitter, and Statistical Outlier Removal (SOR). Rotation involves rotating the point cloud around the $z$-axis by a random angle of $10^{\circ}$, whereas Rotation-3D \purple{extends it to} all three axes. Scaling and Translation are implemented following uniform distributions $\mathcal{U}(0.95, 1.05)$ and $\mathcal{U}([-0.05, 0.05]^3)$, respectively. The dropout technique randomly eliminates 0\% to 20\% of points, and Jitter introduces a point-wise Gaussian noise $\mathcal{N}(0, 0.02)$. Statistical Outlier Removal identifies and removes outlier points based on local point density, considering a point $p$ an outlier if the number of points within its neighborhood $B(p, 0.1)$ is less than 50. To thoroughly assess the sensitivity of various attacks to each data augmentation, we apply them individually as presented in Tab.\ref{tab:def}. The findings demonstrate that our method remains robust against prevalent data augmentations, whereas PointBA-I shows vulnerability to SOR, and rotation-based augmentations tend to overlap PointBA-O.

\subsubsection{Resistance to Expert Defensive Techniques} \label{pointcrt_section}

PointCRT \cite{pointcrt} is the only existing work that specifically targets the backdoor attacks on 3D point clouds. 
\bluesection{
It considers a user-informed scenario where the users master the method of trigger implanting and access the one-hot prediction of the hijacked model. The users collect a benign subset from ModelNet10 and generate the polluted version with the informed attack. Then, they encode the point clouds according to their resilience against 15 distinct corruptions and train a nonlinear binary classifier to discern the compromised data.
}
The F1 score (F1) and Area Under Curve (AUC) are used to assess whether PointCRT can identify the attack, with lower scores indicating stronger penetration ability. Tab.\ref{pointcrt} shows the performance of PointCRT, among which iBA's performance is second only to PointBA-O. \bluesection{By adjusting the attack intensity, PointCRT's performance increases for stronger PointBA-O and decreases for weaker iBA. Thus, further improving the imperceptibility can enhance iBA's resistance to PointCRT.}

Considering that current backdoor defenses for 3D point clouds are very limited, we also introduce two adversarial defenses in Tab.\ref{tab:def}. \cite{lpf} observed that the patterns of adversarial attacks typically reside in the high-frequency portion of the spherical harmonic domain and consequently designed a low-pass filter (LPF) to eliminate the attack. \cite{dup} proposes to denoise and upsample (DUP) the data. Results in Tab.\ref{tab:def} indicate that iBA can also resist these countermeasures.

\subsubsection{Resistance to Gradient-Based Salience Detection}
Backdoor patterns are often regarded as \textit{salient points} due to their significant impact on decision-making. Thus, we can reveal some trigger patterns by detecting the salient points. We apply a gradient-based method proposed in \cite{salience}, which perturbs a point cloud $X$ by moving a point $x_i \in X$ towards the center.

If reducing $r_i=\|x_i\|$ results in an increase in loss $\mathcal{L}$, it indicates that $x_i$ positively contributes to the correct inference. Define $\rho_i = r^{-\alpha}$ ($\alpha>0$ is a scaling factor). Since $\rho_i$ and $r_i$ are inversely proportional, the $\partial \mathcal{L} / \partial \rho_i$ can be directly defined as the significance value $s_i$.

We choose $\alpha=1$ and compare our method with PointBA-I in Fig.\ref{saliency}. The top 2\% \purple{most significant} points are marked in red. As expected, the ball pattern of PointBA-I is well exposed, whereas the trigger of iBA, being dispersed globally, is difficult to locate. Similar results are also presented in \cite{mba}.

\subsubsection{Traits Under 3D Grad-CAM} \label{cam-section}

Gradient Class Activation Map (Grad-CAM)\cite{gradcam} also utilizes a model's gradients to spotlight critical image areas for class identification, offering visual insights into the model's decision-making process and enhancing our understanding of its behavior. Unlike the previous section, where \cite{salience} considers the relationship between \textit{point} perturbation and \textit{loss}, Grad-CAM provides a more fine-grained analysis of the relationship between \textit{feature} perturbation and the confidence of a specific \textit{class}. Therefore, Grad-CAM can more profoundly reveal the backdoor pattern and the characteristics of the hijacked model.

\begin{figure*}[!htbp]
\centering
\includegraphics[width=1.0\linewidth]{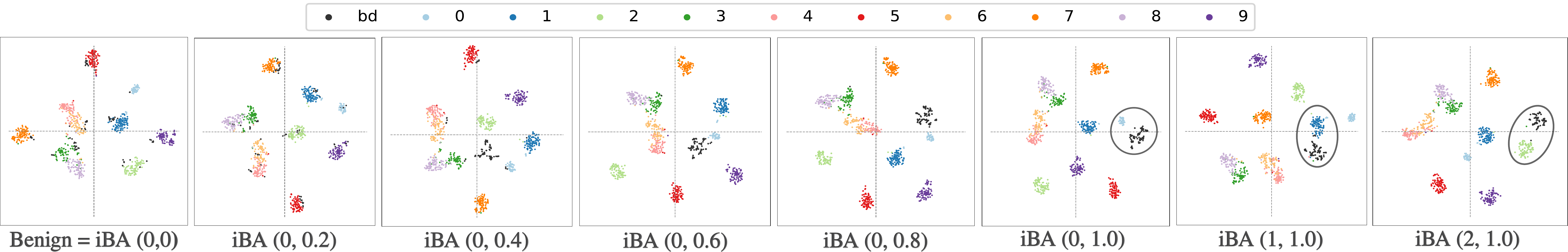}
\caption{DGCNN's feature visualization on ModelNet10 with t-SNE dimensionality reduction\cite{tsne}. "bd" denotes backdoor samples. "iBA (0, 0.2)" represents iBA targeting $0$-th class with intensity $t=0.2$. Subplots 1-6 demonstrate that backdoor samples increasingly deviate and cluster as the intensity $t$ rises. Subplots 6-8 show that the target label induces backdoor samples \bluesection{to} approach and cluster around the indigenous ones (marked with gray circles).
All the subplots share a\purple{n} $[-55, 55]^2$ axis range.} 
\label{tsne}
\end{figure*}

\begin{figure}[!htbp]
\centering
\includegraphics[width=0.5\linewidth]{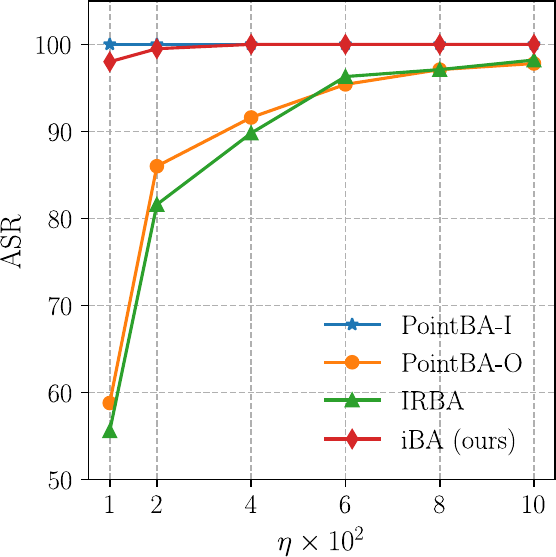}
\caption{The effect of the poisoning rate $\eta$ towards DGCNN on ModelNet10. Our method maintains a high ASR even at a very low poisoning rate (1\%), while the ASRs of PointBA-O and IRBA significantly decrease.} 
\label{pr}
\end{figure}

We implement Grad-CAM for 3D point clouds as follows: Assume that the point cloud $X \in \mathbb{R}^{N \times 3}$ is encoded by a DGCNN backbone, resulting in a point-wise feature $A_X \in \mathbb{R}^{N \times c}$, and a subsequent classifier $cls: A_X \mapsto \hat{y}$ maps these features to a soft prediction. Here, 
\begin{equation} \label{gradcam1}
    w^{i}_{j,k} =\frac{\partial \hat{y}_i}{\partial A_{X,j,k}}, \;
    w^{i}_{k} = \sum_{j} w^{i}_{j,k},
\end{equation}
represent the response level of each spatial location $j$ and channel $k$ within the feature $A_X$ to the $i$-th class respectively. We obtain channel-wise attention $w^{\purple{i}}_k \in \mathbb{R}^c$ by summing across the spatial dimension and \bluesection{utilizing} it to reweight the feature map $A_X$. Then, a $\operatorname{ReLU}$ is attached to retain those channels that have a positive response to the $i$-th class. Finally, we have the saliency map $X^{i}_{s} \in  \mathbb{R}^{N}$:
\begin{equation} \label{gradcam2}
    X^{i}_{s} = \operatorname {ReLU}(\sum_{k} w^{i}_{k}A_{X,:,k}).
\end{equation}
where $A_{X,:,k}$ denotes the slice of $A_X$ along channel $k$.

Multiple experiments are conducted involving various data $X$, models $f$, and reference labels $y_i$ in Fig.\ref{grad-cam}. For instance, Exp.3 utilizes a benign model to identify PointBA-I data and derive partial derivatives for each sample's initial category. Exp.0 serves as \purple{the} blank control group, reflecting \bluesection{benign} significant areas. \bluesection{Exp.1 demonstrates that the ball pattern is easily recognized by the 3D Grad-CAM, which can relay SOR when the cluster lies near the object. Exp.2-3 shows that iBA and PointBA-O are difficult to locate due to their dispersed patterns.} The hijacked model in Exp.4 results in the inability to normally display significant areas. Exp.5 reveals that the trigger is only exposed when we take the partial derivative of the target class $y_t$. 
In Exp.6, the bed's original category and the target label overlap, and the 3D Grad-CAM fully highlights the trigger while ignoring the original patterns, indicating that the trigger is a very strong feature\cite{rethinking}. 

\begin{table}
\begin{center}
\caption{Effect of the target label }
\label{tab:target_effect}
\renewcommand \arraystretch{1.2}
\resizebox{1.0\linewidth}{!}{
\begin{tabular}{c|cc|cc|cc}
\hline
\multirow{2}{*}{Datasets}  & \multicolumn{2}{c|}{$y_t=0$}          & \multicolumn{2}{c|}{$y_t=1$}        & \multicolumn{2}{c}{$y_t=2$}       \\ \cline{2-7} 
                         & \multicolumn{1}{c|}{ACC$\uparrow$}  & ASR$\uparrow$  & \multicolumn{1}{c|}{ACC$\uparrow$}  & ASR$\uparrow$  & \multicolumn{1}{c|}{ACC$\uparrow$}  & ASR$\uparrow$  \\ \hline \hline
\multirow{1}{*}{ModelNet40}     & \multicolumn{1}{c|}{90.4} & 99.4 & \multicolumn{1}{c|}{90.7} & 99.5 & \multicolumn{1}{c|}{90.5}  & 99.5 \\ \hline
\multirow{1}{*}{ModelNet10}     & \multicolumn{1}{c|}{93.2} & 99.5  & \multicolumn{1}{c|}{93.3} & 100  & \multicolumn{1}{c|}{93.3} & 99.8  \\ \hline
\multirow{1}{*}{ShapeNetPart}   & \multicolumn{1}{c|}{98.7} & 99.8  & \multicolumn{1}{c|}{98.7} & 99.1  & \multicolumn{1}{c|}{98.9} & 99.8  \\ \hline
\end{tabular}
}
\end{center}
\end{table}

We also conduct a counterfactual study in Exp.7 by inversing the gradients and acquiring $X^{i}_{-s}$, which shows the region competing with $X^{i}_{s}$. In our task, the trigger pattern and the original feature form a \textit{competitive} relationship. For target class $y_t=$ "bed", $X^{i}_{-s}$ exhibits a reverse form of $X^{i}_{s}$ in Exp.6, indicating that $X^{i}_{-s}$ can release the original features suppressed by the trigger (blue box in Fig.\ref{grad-cam}). For other non-target classes, the highlighted regions almost coincide with those activated by the target class $y_t$ shown in Exp.1 (yellow boxes in Fig.\ref{grad-cam}), which completely aligns with our reasoning about the competitive region.

The 3D Grad-CAM also helps us understand the impact of the target label, which we will discuss in Sec.\ref{target} below.

\subsection{Ablation Studies}

\subsubsection{Effect of the Poisoning Rate}
The poisoning rate $\eta$ determines the trade-off between attack concealment and effectiveness. The ideal attack should achieve a high ASR with a small $\eta$. In Fig.\ref{pr}, we explore the trend of ASR with increasing $\eta$ in different attacks. The results show that our transformation-based iBA maintains a high ASR even at a very low $\eta$, benchmarking against interaction-based methods. In contrast, other transformation-based methods suffer from significant performance degradation.

\subsubsection{Effect of the Target Category} \label{target}
We and our pioneer\purple{s}\cite{pointba}\cite{irba} choose several representative target categories to save on experimental expenses. One may wonder whether the results has generality. \cite{irba} provides empirical results and speculates that the target label has little impact on IRBA. However, the principle behind this phenomenon remains elusive.
To remedy this gap, we combine the latest theoretical research\cite{rethinking} as well as experiments from multiple perspectives in Tab.\ref{tab:target_effect}, Fig.\ref{tsne}, and Fig.\ref{grad-cam} to further discuss the impact of target label. Tab.\ref{tab:target_effect} presents the performance of our method when we select other target categories, with the victim model being DGCNN. Switching target categories does not significantly affect the performance. 

Some works might help us understand this phenomenon. \cite{neuralcleanse} suggests that the essence of backdoor attacks lies in establishing shortcuts between backdoor samples and target categories in the victim network. Namely, the trigger pattern bypasses the pathway of the source image in the inference graph, activating a shorter, alternative pathway.
Meanwhile, \cite{rethinking} assumes that the trigger often acts as the \textit{strongest} feature, suppressing the original features of the source images. Removing it leads to an effective defensive performance. The 3D Grad-CAM reveals that the trigger pattern has gained an overwhelming advantage in the competition with the original ones (see the blue and yellow boxes in Fig.\ref{grad-cam}), which corroborates the assumptions of  \cite{neuralcleanse} and \cite{rethinking}. 

\begin{figure} 
    \centering
  \subfloat[Imperceptibility $w.r.t. \; t$ and $N_l$\label{smooth_config_1}]{%
       \includegraphics[width=0.52\linewidth]{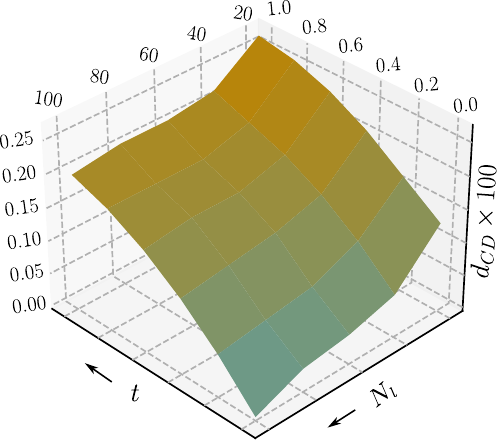}}
    \hfill
  \subfloat[ASR $w.r.t. \; t$\label{smooth_config_2}]{%
        \includegraphics[width=0.43\linewidth]{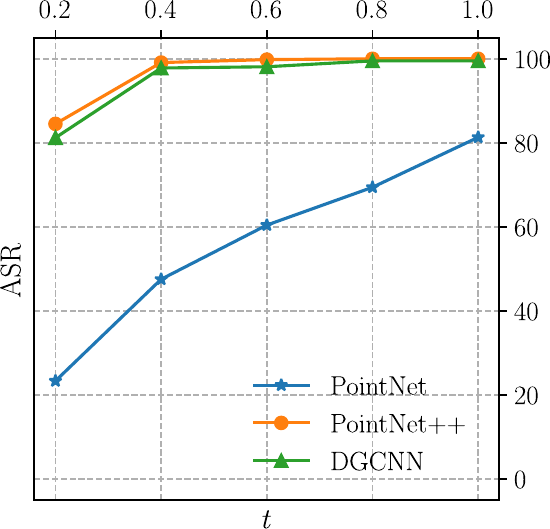}}
  \caption{The impact of trigger smoothing parameters on the imperceptibility and ASR. (a) Initially, at $t=0$, the curve shows the influence of the spherical harmonics expansion order $N_l$ on the reconstruction error, which becomes negligible when $N_l=100$. Fixing $N_l=100$, the resulting curve shows that, as $t$ increases, the imperceptibility will drop while ASR will rise (b).} 
  \label{smooth_config} 
\end{figure}

Fig.\ref{tsne} visualizes the feature space of the ModelNet10 test set with t-SNE\cite{tsne}. Regardless of \purple{the} source classes, iBA samples always manage to distance themselves from the source samples and form a new cluster near the indigenous ones. Of course, separability in feature space is not a necessary condition for a successful attack\cite{rethinking}\cite{topological}. iBA with a low attack intensity ($t \leq 0.2$) produces imperceptible alterations in the feature space. However, it still achieves a considerable ASR (see Fig.\ref{smooth_config_2}, Sec.\ref{smooth_config_para}), devoid of feature separability. It is foreseeable that iBA \purple{with lower intensities} will be able to resist future backdoor defenses based on the feature separation assumption.

\begin{table}
\begin{center}
\caption{Trigger imperceptibility in different configurations. The synergistic effect of two losses generates the most imperceptible trigger} 
\label{tab:dist_ablation}
\renewcommand \arraystretch{1.2}
\resizebox{0.9\linewidth}{!}{
\begin{tabular}{c|c|c|c}
\hline
Methods               & CD $\times$ 100 $\downarrow$ & WD $\times$ 0.1$\downarrow$ & HD$\downarrow$ \\ \hline \hline
iBA                & 0.20            & 0.81             & 0.14          \\ \hline
iBA (w/o $d_{SWD}$) & 0.21            & 1.41             & 0.13          \\ \hline
iBA (w/o $d_{CD}$)  & 0.32            & 0.83             & 0.16          \\ \hline
\bluesection{iBA (Diffusion)}         & \bluesection{0.10}            & \bluesection{0.61}             & \bluesection{0.08}          \\ \hline
\end{tabular}
}
\end{center}
\end{table}

\begin{table}
\begin{center}
\caption{Trigger effectiveness in different configurations}
\label{tab:asr_ablation}
\renewcommand \arraystretch{1.2}
\resizebox{1.0\linewidth}{!}{
\begin{tabular}{c|cc|cc|cc}
\hline
\multirow{2}{*}{Methods}                   & \multicolumn{2}{c|}{PointNet}           & \multicolumn{2}{c|}{PointNet++}          & \multicolumn{2}{c}{DGCNN}       \\ \cline{2-7} 
                                           & \multicolumn{1}{c|}{ACC$\uparrow$}  & ASR$\uparrow$    & \multicolumn{1}{c|}{ACC$\uparrow$}  & ASR$\uparrow$     & \multicolumn{1}{c|}{ACC$\uparrow$}  & ASR$\uparrow$  \\ \hline \hline
\multirow{1}{*}{iBA}                    & \multicolumn{1}{c|}{91.7} & 81.3  & \multicolumn{1}{c|}{93.1} & 100    & \multicolumn{1}{c|}{94.0} & 99.5 \\ \hline
\multirow{1}{*}{iBA (w/o $d_{SWD}$)}     & \multicolumn{1}{c|}{92.0} & 72.6  & \multicolumn{1}{c|}{92.5} & 98.6   & \multicolumn{1}{c|}{93.1} & 99.3  \\ \hline
\multirow{1}{*}{iBA (w/o $d_{CD}$)}      & \multicolumn{1}{c|}{91.9} & 98.0  & \multicolumn{1}{c|}{93.8} & 100    & \multicolumn{1}{c|}{93.8} & 100  \\ \hline
\multirow{1}{*}{\bluesection{iBA (Diffusion)}}        & \multicolumn{1}{c|}{\bluesection{91.7}} & \bluesection{4.3}  & \multicolumn{1}{c|}{\bluesection{93.1}} & \bluesection{94.3}    & \multicolumn{1}{c|}{\bluesection{93.9}} & \bluesection{97.9}  \\ \hline
\end{tabular}
}
\end{center}
\end{table}

\subsubsection{Effect of the Trigger Configuration} \label{AE_achitecture}
Below, we discuss the impact of different configurations on the \bluesection{trigger generation}.  Tab.\ref{tab:dist_ablation} presents the attack imperceptibility on ModelNet10. The synergy between $d_{CD}$ and $d_{SWD}$ makes the final iBA very stealthy across three metrics. The absence of either $d_{CD}$ or $d_{SWD}$ can lead to more conspicuous triggers under the respective metrics. Tab.\ref{tab:asr_ablation}, on the other hand, measures the effectiveness of different \bluesection{variants}. \purple{It shows that the different configurations in these folding-based variants result in distinct performances. The absence of $d_{SWD}$ diminishes the attack effectiveness against PointNet, whereas excluding $d_{CD}$ significantly degrades the reconstruction quality.} In addition, the trigger generated with a more advanced diffusion-based AE exhibits greatly improved imperceptibility but loses \purple{the} effectiveness.

\subsubsection{Effect of the Smoothing Configuration} \label{smooth_config_para}
The trigger smoothing module based on SHT has two crucial parameters: the attack intensity $t$ and the maximum order $N_l$. We demonstrate their influence on ModelNet10 in Fig.\ref{smooth_config_1}. On the one hand, the deviation of the backdoor data (measured in CD) rises steadily as $t$ increases. On the other hand, the truncation error nearly vanishes when $t=0$ and $N_l=100$, verifying the rationality of our choice.  Fig.\ref{smooth_config_2} shows the impact of attack intensity $t$ on the ASR. Both PointNet++ and DGCNN are extremely sensitive to iBA since $t=0.4$ has already ensured the effectiveness. The imperceptibility of the trigger is significantly enhanced while maintaining effectiveness, which once again confirms the importance of trigger smoothing.

\bluesection{
 
\section{Trigger Pattern Analysis} \label{pattern}

This section will \purple{investigate} the specific patterns generated by the folding-based AE from the perspective of frequency. 
We first transform the point cloud to the frequency domain with graph Fourier transform (GFT) $\phi_{GFT}:\mathbb{R}^{n \times 3} \rightarrow \mathbb{R}^{n}$ (we average the output signal in the spatial dimension for simplification), where $n=1024$ and the graph Laplacian is constructed with 10-nearest neighbors. Specifically, given benign data $X$ and a TIF $\mathcal{G}$, we have the residual signal:
\begin{gather} \label{residual_GFT}
\lambda^{res}=\phi_{GFT}(X)-\phi_{GFT}(\mathcal{G}(X)),
\end{gather}
where $\lambda^{res}=(\lambda_{1}^{res}, \lambda_{2}^{res}, \cdots, \lambda_{1024}^{res})$. We further quantify and normalize $\lambda^{res}$ by:
\begin{gather} \label{quantify_residual_GFT}
\lambda_{R_{k}}^{res}=\frac{\sum_{i \in R_k}|\lambda_{i}^{res}|}{\sum_{i \in R}|\lambda_{i}^{res}|},
\end{gather}
where $k=1,2,\cdots, 6$ and $R=R_1 \cup R_2 \cup \cdots \cup R_6$. $R_1, R_2$ are ultra-low (UL) and low (L) frequency bands that control the global silhouette. $R_3, R_4$ are lower-mid (LM) and higher-mid (HM) bands that determine the local geometries. $R_5, R_6$ are high (H) and ultra-high (UH) bands representing details and noises. $1, 8, 32, 128, 256, 512, 1024$ are the corresponding endpoints of six bands, i.e., $R_1=\{1, 2, \cdots, 8\}$ and so on. Quantification can prevent high-frequency information from being overwhelmed by the principal components. Normalization allows us to focus on the characteristics of the patterns rather than the specific attack intensities. 

\begin{figure} 
\captionsetup{labelfont={color=blue},font={color=blue}}
    \centering
  \subfloat[\bluesection{Common triggers}\label{GFT1}]{%
       \includegraphics[width=0.33\linewidth]{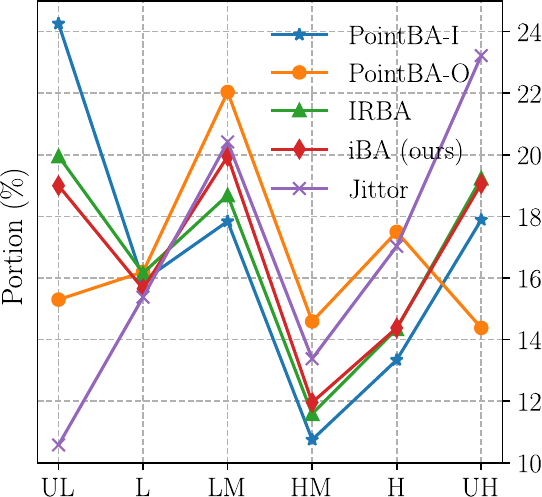}}
    \hfill
  \subfloat[\bluesection{Our variants}\label{GFT2}]{%
        \includegraphics[width=0.33\linewidth]{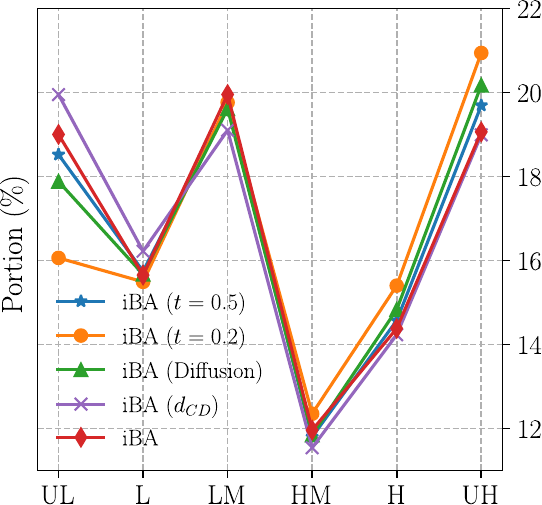}}
    \hfill
  \subfloat[\bluesection{References}\label{GFT3}]{%
        \includegraphics[width=0.33\linewidth]{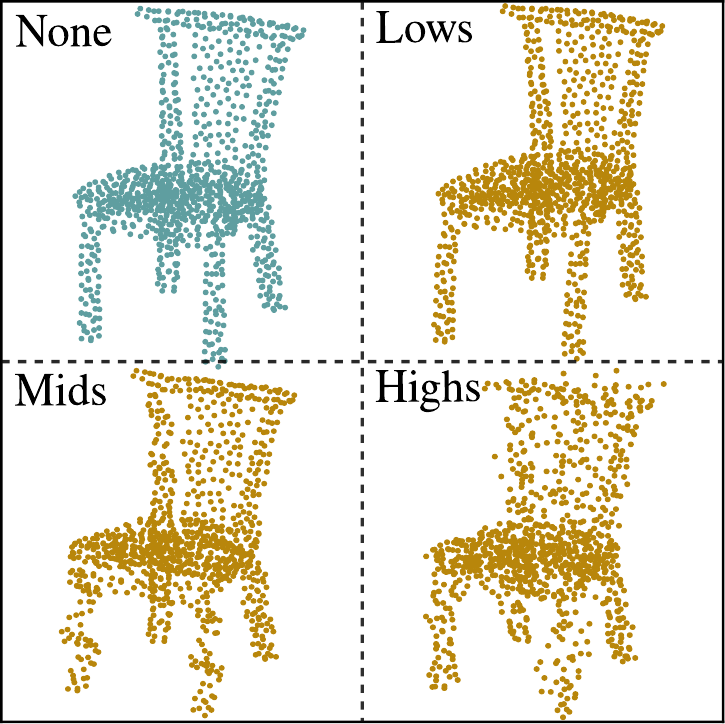}}
  \caption{\bluesection{Trigger pattern analysis on ModelNet10 test set. (a-b) The distribution of trigger patterns in different frequency bands, including ultra-lows (UL), lows (L), lower-mids (LM), higher-mids (HM), highs (H), and ultra-highs (UH). (c) Visualization of geometric patterns caused by perturbations in different frequency bands.}} 
  \label{GFT} 
\end{figure}

Fig.\ref{GFT} reports the averaged results on ModelNet10. Fig.\ref{GFT1} compares iBA with other attacks. We also introduce jitter, a high-frequency point-wise perturbation, for reference. PointBA-I and PointBA-O focus on the low and mid frequencies, respectively. iBA and IRBA both disturb a wide range of frequency bands. However, IRBA emphasizes the low ones while we prioritize the mid ones, which explains 
our immunity to denoisers like DUP and LPF. Fig.\ref{GFT2} illustrates the characteristics of multiple variants. The low-frequency parts increase significantly without the global constraints from $d_{SWD}$. Variants with a lower $t$ or diffusion model tend to shift the attack strength towards high frequencies, accounting for their defective ASR on PointNet, a model that focuses less on high-frequency details.

\section{Potential Defenses} \label{potential_defense}
In Sec.\ref{pointcrt_section}, PointCRT considers a user-informed scenario and achieves impressive performance. The user-informed scenario assumes that the users anticipate the risk of being attacked and \purple{master the skill of generating} specific backdoor patterns. Like creating an antivenom from the venom itself, this scenario will become increasingly important as the users' security awareness rises and pose greater challenges to the attacker. Thus, we now discuss the possible defenses in the user-informed scenario.

Since the correctness and accuracy of the information informed can vary in real applications, we consider the following five cases:
\begin{itemize}
    \item C$_0$: the users realize the attack but mistakenly regard it as IRBA, which serves as the blank control group to demonstrate the importance of information correctness. 
    \item C$_1$: the users know that the attacker utilizes an AE but mistakenly recognize it as the diffusion-based one. 
    \item C$_2$: the users get the exact folding-based trigger injector but train it with only Chamfer loss.
    \item C$_3$: the users almost crack the attacker's method but set the smoothing parameter $t=0.5$ instead of $t=1.0$.
    \item C$_4$: the users completely crack the attacker's iBA trigger ($t=1.0$).
\end{itemize}
Then, we will discuss two user-informed defense strategies in these cases. In addition, two unsupervised out-of-distribution detection techniques are evaluated too.
Unless otherwise mentioned, all the defenses are tailored to the default iBA ($t=1.0$).

\subsection{Dataset Reconstruction with User-informed Triggers}
Assume that the training set \purple{that} the users have is 
\begin{gather*}
    \mathcal{D}' = \mathcal{D}_c  \cup \mathcal{D}'_{b} = \{(X_{c_i}, y_{c_i})\}_{i=1}^{N_c} \cup \{(\mathcal{G}_{A}(X_{b_j}), y_t)\}_{j=1}^{N_b}, 
\end{gather*}
where $\mathcal{D}_c$ and $\mathcal{D}'_{b}$ are the benign and backdoor subsets containing $N_c$ and $N_b$ samples indexed with $c_i$ and $b_j$, respectively. $\mathcal{G}_{A}$ is the attacker's TIF with target label $y_t$. $\mathcal{G}_{U}$ is the user-informed trigger, and the users now reconstruct the entire $\mathcal{D}'$ with labels reserved:
\begin{gather*} 
    \mathcal{G}_{U}(\mathcal{D}') = \mathcal{G}_{U}(\mathcal{D}_c) \cup \mathcal{G}_{U}(\mathcal{D}'_{b}), \\
    \mathcal{G}_{U}(\mathcal{D}_c) :=  \{(\mathcal{G}_{U}(X_{c_i}), y_{c_i})\}_{i=1}^{N_c}, \\
    \mathcal{G}_{U}(\mathcal{D}'_{b}) := \{(\mathcal{G}_{U}(\mathcal{G}_{A}(X_{b_j})), y_t)\}_{j=1}^{N_b}. 
\end{gather*}

Two defensive mechanisms will work in the dataset reconstruction process. First, the data in $\mathcal{G}_{U}(\mathcal{D}'_{b})$ are polluted twice, and the original trigger $\mathcal{G}_{A}$ can be damaged by the succeeding $\mathcal{G}_{U}$. Second, when $\mathcal{G}_{U} \approx \mathcal{G}_{A}$, the correctly-labeled $\mathcal{G}_{U}(\mathcal{D}_c) \approx \mathcal{G}_{A}(\mathcal{D}_c)=\{(\mathcal{G}_{A}(X_{c_i}), y_{c_i})\}_{i=1}^{N_c}$ will guide the model to infer correctly on the backdoor data. We call the two mechanisms \textit{trigger damage} and \textit{benign guidance}, respectively. For instance, benign guidance plays a crucial role in defending PointBA-O with rotation augmentation in Tab.\ref{tab:def}. 

\begin{table}
\captionsetup{labelfont={color=blue},font={color=blue}}
\begin{center}
\caption{\bluesection{User-informed defense by data reconstruction on ModelNet10}} 
\label{user_informed_recon}
\renewcommand \arraystretch{1.2}
\resizebox{0.8\linewidth}{!}{
\begin{tabular}{c|c|c|c|c|c}
\hline
                & C$_0$  & C$_1$ & C$_2$ & C$_3$ & C$_4$ \\ \hline \hline
ACC$\uparrow$   & 91.6  & 93.4  & 71.1 & 88.9  & 76.3  \\ \hline
ASR$\uparrow$   & 96.9  & 21.6  & 5.6  & 18.0  & 2.1  \\ \hline

\end{tabular}
}
\end{center}
\end{table}

We consider all five cases above, and the experimental results are shown in Tab.\ref{user_informed_recon}.
Here, we reconstruct all data offline before training rather than augmenting online, as in Tab.\ref{tab:def} since some trigger generation processes can be time-consuming and online augmentation is impractical. When the users incorrectly use IRBA to reconstruct data, the attack is not effectively mitigated \purple{while} all variants of iBA achieve satisfactory results.

\begin{table}
\captionsetup{labelfont={color=blue},font={color=blue}}
\begin{center}
\caption{\bluesection{User-informed defense by data reconstruction with iBA (Diffusion) on ModelNet10}} 
\label{user_informed_recon2}
\renewcommand \arraystretch{1.2}
\resizebox{0.9\linewidth}{!}{
\begin{tabular}{c|c|c|c|c}
\hline
                  & PointBA-I  & PointBA-O  & IRBA & iBA \\ \hline \hline
ACC$\uparrow$     & 93.6       & 93.3       & 93.8 & 93.4  \\ \hline
ASR$\uparrow$     & 57.8       & 2.8        & 6.5  & 21.6  \\ \hline

\end{tabular}
}
\end{center}
\end{table}

However, iBA (Diffusion) achieves better-than-expected performance even though it looks different from the folding-based ones. Both of the mechanisms discussed above may lead to this phenomenon. First, iBA (Diffusion) significantly damages the folding-based patterns. Second, the diffusion-based trigger and folding-based trigger might share some unexplored similarities, which activates the benign guidance mechanism. 
We can currently verify the first one by applying iBA (Diffusion) to other attacks in Tab.\ref{user_informed_recon2}. All of them endure a great ASR declination, indicating the involvement of trigger damage. Since the diffusion model is trained on benign data, the domain gap leads to a noisier reconstruction \purple{on} backdoor data, leading to severe trigger damage.

\subsection{Supervised Classification in User-informed Scenarios}
PointCRT considers a black-box scenario where the users can only access the final hard prediction from the backdoor model. However, in practice, the users may train a surrogate model with the polluted dataset and acquire a compromised deep representation. As Fig.\ref{tsne} depicts, the backdoor and target features are not completely mingled together, suggesting that the compromised deep representation may be discriminative for backdoor recognition. To put it another way, the users can still seek help from traditional point cloud features when the compromised deep features are not trustworthy.

\begin{table}
\captionsetup{labelfont={color=blue},font={color=blue}}
\begin{center}
\caption{\bluesection{User-informed defense using the supervised classifier on ModelNet10 with F1-score reported, DGCNN-C is the compromised DGCNN representation}} 
\label{tab:supervised}
\renewcommand \arraystretch{1.3}
\resizebox{1.0\linewidth}{!}{
\begin{tabular}{c|c|c|c|c|c}
\hline
\multicolumn{1}{l|}{Features}      & C$_0$     & C$_1$     & C$_2$     & C$_3$   & C$_4$     \\ \hline \hline
\multirow{1}{*}{DGCNN-C}           &  0.746    & 0.836      & 0.840   & 0.843     & 0.861 \\ \cline{1-6} 
\multirow{1}{*}{FPFH}              &  0.761    & 0.818      & 0.815   & 0.877     & 0.902       \\ \cline{1-6} 

\end{tabular}
}
\end{center}
\end{table}

We explore both representations for learning a backdoor recognizer. Specifically, we utilize DGCNN as the surrogate model for compromised features (DGCNN-C) and \purple{we adopt} Fast Point Feature Histogram (FPFH)\cite{fpfh} as the traditional representation. FPFH is a computationally efficient descriptor that characterizes the local geometric properties of 3D point clouds, providing robust and distinctive features for object recognition. For the binary classifier, we employ AdaBoost\cite{adaboost} with weak learners based on the decision tree. The users are assumed to acquire 10 benign samples (one for each class in ModelNet10). They further construct 10 backdoor ones with the informed attack to train the classifier. 

\begin{table}
\captionsetup{labelfont={color=blue},font={color=blue}}
\begin{center}
\caption{\bluesection{Detection rate \purple{(\%)} with unsupervised anomaly detection methods Isolated Forest (IF) and One-class SVM (OC-SVM), $\tau$ is the anomalous margin}} 
\label{tab:unsupervise}
\renewcommand \arraystretch{1.3}
\resizebox{1.0\linewidth}{!}{
\begin{tabular}{c|c|c|c|c|c}
\hline
\multicolumn{1}{l|}{Features}  & Methods          &  $\tau=0.05$ & $\tau=0.1$  & $\tau=0.15$ & $\tau=0.2$ \\ \hline \hline
\multirow{2}{*}{DGCNN-C}       & IF               &  13.3       & 26.7       & 35.6       & 42.3    \\ \cline{2-6} 
                               & OC-SVM           &  21.2       & 32.2       & 44.4       & 55.6       \\ \cline{1-6} 

\multirow{2}{*}{FPFH}          & IF               &  15.6       & 44.4       & 70.0       & 80.0       \\ \cline{2-6} 
                               & OC-SVM           &  26.7       & 55.0       & 77.8       & 91.1       \\ \cline{1-6} 

\end{tabular}
}
\end{center}
\end{table}

We test the classifier on ModelNet10 in Tab.\ref{tab:supervised}. With only 20 training samples, all variants of iBA demonstrate considerable recognition accuracy with both DGCNN-C and FPFH features. The supervised classifier trained on IRBA data can roughly recognize iBA as well, indicating its transferability.
A natural question is: how can we improve our attack in the future to resist the supervised classifiers? The key lies in degrading the discrimination from the representation. Invisible attacks in the feature space\cite{featureinvisible} will disable the compromised deep representation while preserving local geometric details as much as possible might \purple{blind the} traditional representations.

\subsection{Unsupervised Anomaly Detection}
The discussion above reveals that \purple{both} DGCNN-C and FPFH features can recognize backdoor data. Can they still be helpful when the users anticipate the attack but no extra information is available? We frame this as an application of unsupervised anomaly detection. We test Isolation Forest\cite{if} and One-class SVM (with RBF kernel)\cite{oneclasssvm} based on the two representations, respectively. Concretely, we randomly trigger 5\% \purple{of the} data (5 for each non-target class, \purple{totaling 45 samples}), \purple{as well as} set the anomalous margin $\tau$ to 0.05, 0.1, 0.15, and 0.2. $\tau=0.2$ means $20\%$ data are regarded poisoned, which should be carefully chosen to trade off the detection rate and the benign data sacrifice. We experiment twice and present the average detection rates in Tab.\ref{tab:unsupervise}.
One-class SVM and FPFH show superior performance. 

\blue{
\subsection{Discussion on Future Defenses}
The experiments demonstrate that each of the three strategies possesses unique characteristics and can be effective in appropriate scenarios, providing valuable insights for future defenses against AE-based attacks like iBA.

High-performance pretrained AEs are ideal tools for dataset reconstruction. They can effectively damage the iBA trigger while minimally impacting the original features. However, they are not always available. Users can instead train AEs from scratch using the compromised dataset. The extreme imbalance in the dataset may still result in poorer reconstruction of backdoor data, thereby mitigating the attack. Therefore, future research could explore the performance of these compromised AEs.

When users have access to benign reference samples, supervised detectors should be prioritized. A simple boosting classifier can already recognize iBA data based on the compromised deep features. Experiments show that supervised detectors can even transfer to unseen attacks, suggesting that exploring this potential in future studies could be beneficial.

Unsupervised anomaly detectors heavily rely on feature representation. Therefore, the key to successfully applying this strategy is to identify more discriminative features for iBA-like data. Enhancing features would also benefit supervised methods.
}

\section{Potential of physical attack} \label{physical_potential}

In this section, we explore the potential for deploying iBA in real-world scenarios. Similar to \cite{irba}, we utilize 3D printing technology \cite{3dprint} to fabricate a compromised physical object, which will trigger the attack upon scanning.
Given the variability of scanners and scanning environments, the resulting point clouds can differ significantly. Consequently, only attacks that alter the underlying surface of the point cloud are viable for practical deployment, and non-shape-altering point movements are ineffective\cite{hide}. As analyzed in Sec.\ref{pattern}, our attack disrupts low- and mid-frequency bands, inducing real shape alterations that persist across different scanning methods. This indicates our potential for physical deployment.

To further investigate this potential, we conduct a simulated re-scanning experiment. We begin by removing outliers from the compromised data and converting them into meshes using the Ball-pivoting Algorithm\cite{ball}. We then simulate two types of scanners: a handheld short-range 3D scanner and a vehicle-mounted Light Detection and Ranging (LiDAR) system.
The handheld 3D scanner, ideal for detailed scanning of small objects, typically produces uniform and complete 3D point clouds. We simulate this by uniformly sampling 1024 points from the reconstructed mesh. In contrast, LiDAR, used for large-scale outdoor scene perception in autonomous vehicles, generates uneven and incomplete point clouds. We replicate this by sampling 10 evenly spaced layers along the height dimension to mimic its multi-line laser scanning pattern.

\begin{figure}[!htbp]
\centering
\includegraphics[width=0.9\linewidth]{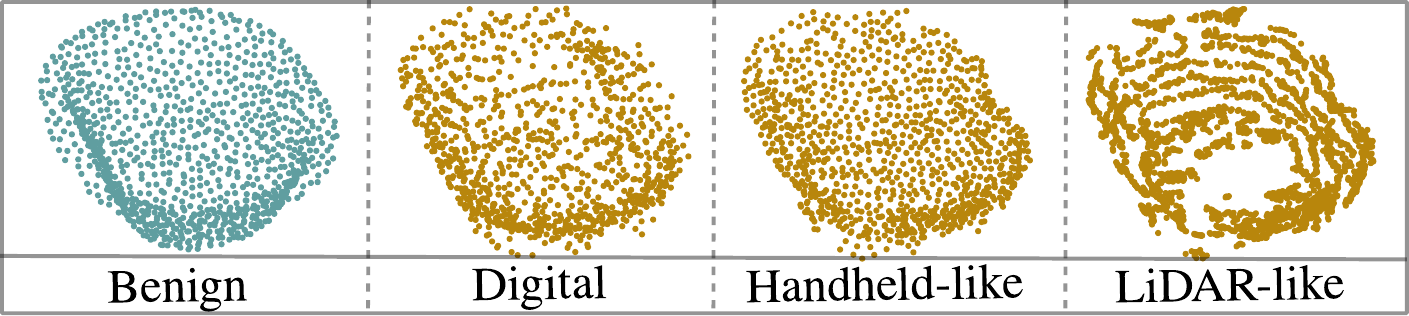}
\caption{\bluesection{Display of the simulated scanning. We reconstruct a mesh from the backdoor data and resample from it to simulate handheld scanners and LiDARs, respectively.}} 
\label{physical}
\end{figure}

The experimental results are illustrated in Tab.\ref{physical_test}. The handheld-like data retains the backdoor latent surface while the point distribution nearly returns to benign, purely stripping the non-shape-altering point movements and thus reducing the ASR. However, when LiDAR-like scanning causes a tremendous domain gap, ASR gains the upper hand again since backdoor data is more robust to corruption\cite{pointcrt}.

It is important to note that these experiments are \purple{just} preliminary explorations, evaluating whether our attack remains effective under non-shape-altering point movements. There is a gap between our simulation and reality (e.g., real LiDAR point clouds exhibit self-occlusion patterns). Therefore, iBA's physical attack capability requires further testing in real-world scenarios.

\begin{table}
\captionsetup{labelfont={color=blue},font={color=blue}}
\begin{center}
\caption{\bluesection{Simulations of physical deployment of our attack on ModelNet10}} 
\label{physical_test}
\renewcommand \arraystretch{1.2}
\resizebox{0.9\linewidth}{!}{
\begin{tabular}{c|c|c|c|c}
\hline
                  & Benign  & Digital  & Handheld-like & LiDAR-like  \\ \hline \hline
ACC$\uparrow$     & 94.0    & 94.0     & 85.0          & 41.5          \\ \hline
ASR$\uparrow$     & --      & 99.5     & 66.1          & 89.9          \\ \hline

\end{tabular}
}
\end{center}
\end{table}

}

\section{Limitations and Future Work}

\bluesection{
While introducing $d_{SWD}$ enhances the effectiveness of iBA on PointNet, it still falls short compared to the impressive ASR achieved on other victim models. Although analysis in the frequency domain offers an intuitive explanation, the underlying principles require further investigation. Factors like geometry, topology, intensity, and network structure should be decoupled to isolate the key components. This will also deepen our understanding of the heterogeneity of 3D DNNs.

The folding-based variants are limited in depicting geometry details such as the legs of chairs and tables. The diffusion-based variant fails to attack PointNet. However, its \purple{remarkable} reconstruction ability presents significant opportunities for future improvements. For example, the attacker might strategically increase its error through latent space manipulation, function composition, or loss induction. LIRA\cite{lira} and AdvDoor\cite{advdoor} also provide insights for stronger 3D learning-based triggers with the development of 3D adversarial attacks. 

In addition to 3D printing, more physical attacks on 3D point clouds are worth considering. \purple{Like} glass reflection, fog as a natural medium can interfere with LiDAR scanning, leading to special noise as potential backdoor patterns. Sec.\ref{physical_potential} has shown that handheld 3D scanners and LiDARs can produce specific point cloud distributions. Recall that PIB utilizes the camera fingerprint as a powerful trigger without altering the physical environment, the scanner fingerprint can also serve as an economical, convenient, and imperceptible physical trigger in 3D point clouds.  

}

\section{Conclusion \bluesection{and Social Impact}}

\bluesection{We propose a simple, effective, and imperceptible 3D backdoor attack, iBA, utilizing a folding-based generative model. The designed trigger smoothing module allows for controllable attack intensity and enhances the versatility of iBA. We evaluate iBA’s resistance to various defense methods and discuss the performance of multiple variants. Additionally, we analyze our trigger patterns from a frequency perspective, provide potential countermeasures, and explore the feasibility of physical deployment. We aim to raise awareness of security issues in 3D point cloud applications and encourage further research to strengthen model robustness against backdoor threats.}

\section*{Acknowledgments}
We thank Mr. Gao from Shenzhen International Graduate School, Tsinghua University, for inspiring and assisting us with some of the problems encountered during our research. 
\bluesection{We also appreciate all the anonymous reviewers for their efforts as well as their enlightening and constructive comments.}
We express our gratitude to all the authors of the open-source projects used in this study for generously sharing their work as well.
The responsibility for the content and any remaining errors is exclusively with the authors.

\section{Biography Section}
\begin{IEEEbiography}[{\includegraphics[width=1in,height=1.25in,clip,keepaspectratio]{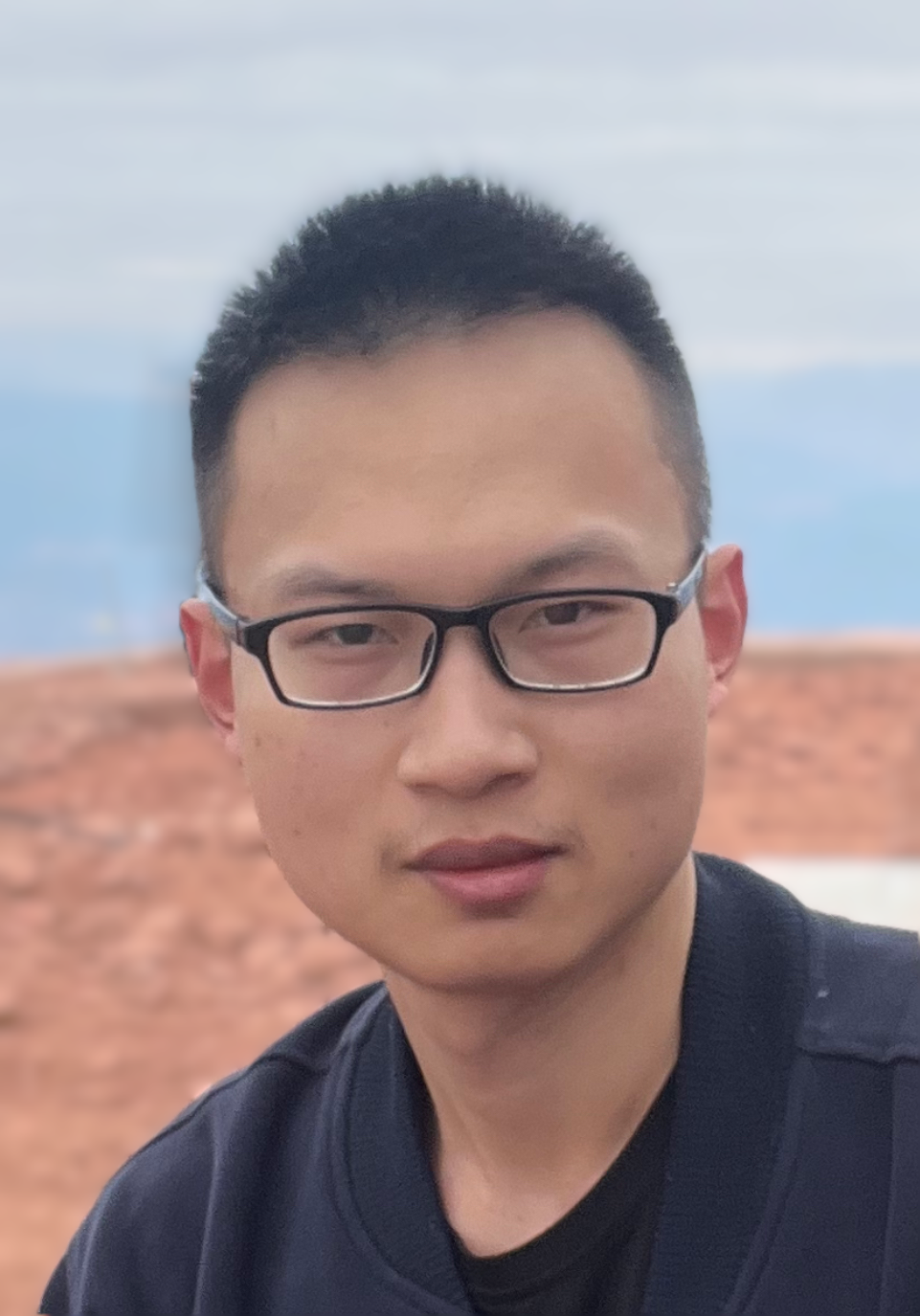}}]{Yuhao Bian}
received the B.A. and M.S. degree from School of Mathematical Sciences, Dalian University of Technology (DLUT). He was a visiting student with Academy of Mathematics and Systems Science, Chinese Academy of Sciences (AMSS, CAS) and an intern engineer in Y-tech, Kuaishou Technology, Beijing, China. He currently pursues the Ph.D. degree with Xiuping Liu. His research interests include adversarial learning and point cloud processing.
\end{IEEEbiography}


\begin{IEEEbiography}[{\includegraphics[width=1in,height=1.25in,clip,keepaspectratio]{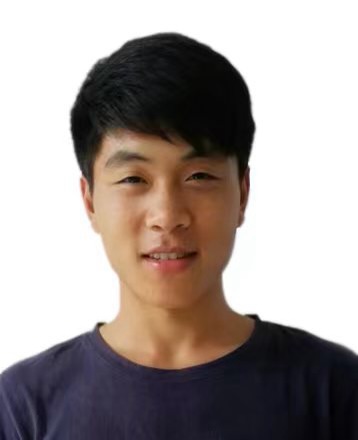}}]{Shengjing Tian}
received the Ph.D. degree from the School of Mathematical Sciences, Dalian University of Technology (DLUT) in 2022. He was a visiting student with the Singapore University of Technology and Design (SUTD) under the supervision of Prof. Jun Liu. He is currently a lecturer and postdoctoral fellow with the China University of Mining and Technology. His research interests include multimodal deep learning, point cloud processing, and video understanding.

\end{IEEEbiography}

\begin{IEEEbiography}[{\includegraphics[width=1in,height=1.25in,clip,keepaspectratio]{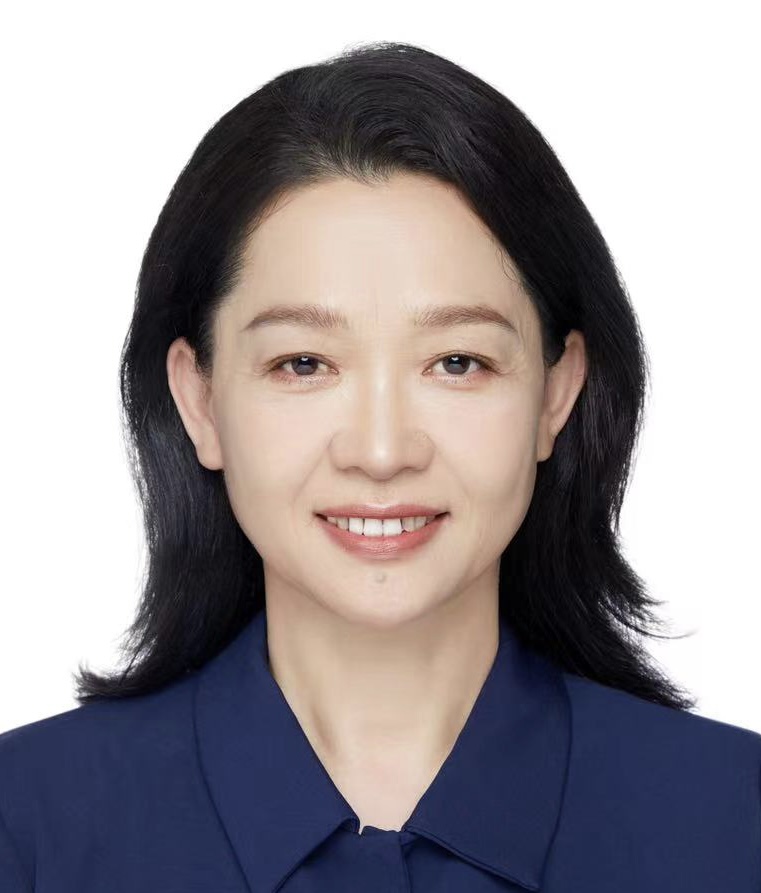}}]{Xiuping Liu}
received the M.S. degree from Jilin University and the Ph.D. degree from the Dalian University of Technology (DLUT). She is currently a Professor with the School of Mathematical Sciences, Dalian University of Technology. She has published many papers in journals and conferences, including Siggraph, ACM Transactions on Graphics, IEEE Transactions on Visualization and Computer Graphics, IEEE Transactions on Image Processing, ECCV, and ICCV. Her research interests are computer vision, computer graphics, and machine learning.

\end{IEEEbiography}

\vfill

\end{document}